\documentclass[12pt]{article}
\usepackage{mathrsfs}

\usepackage{graphicx,setspace,lscape,longtable,epstopdf,xr}
\usepackage{epsfig,graphicx,pdfpages}
\usepackage{bm}
\usepackage{amsmath}
\usepackage{amsfonts}
\usepackage{graphicx,psfrag,epsf}
\usepackage{enumerate}
\usepackage{natbib}
\usepackage{multirow}
\usepackage{url} 
\usepackage{algorithm}
\usepackage{algpseudocode}
\usepackage{graphicx}
\usepackage{chngpage}
\usepackage[T1]{fontenc}
\newcommand{\blind}{0}

\usepackage{booktabs}

\newtheorem{theorem}{Theorem}
\newtheorem{lemma}{Lemma}
\newtheorem{proposition}{Proposition}

\def\one{{\bf 1}}
\def\bA{{\mathcal A}}
\def\ve{\varepsilon}

\def\beq{\begin{equation}}
\def\eeq{\end{equation}}
\def\beqr{\begin{eqnarray}}
\def\eeqr{\end{eqnarray}}
\def\beqrs{\begin{eqnarray*}}
\def\eeqrs{\end{eqnarray*}}
\def\bet{\begin{theorem}}
\def\eet{\end{theorem}}
\def\bel{\begin{lemma}}
\def\eel{\end{lemma}}
\def\bep{\begin{proposition}}
\def\eep{\end{proposition}}
\def\bg{\begin{figure}[tbph]\begin{center}}
\def\eg{\end{center}\end{figure}}

\def\bc{\begin{center}}
\def\ec{\end{center}}

\newtheorem{remark}{Remark}

\RequirePackage[colorlinks,citecolor=blue,urlcolor=blue]{hyperref} 
\def\wt{\widetilde}

\def\wh{\widehat}

\def\ol{\overline }

\def\bB{\mathbf{B}}

\def\mN{\mathcal{N}}

\def\mR{\mathbb{R}}

\def\mV{\mathbb{V}}

\def\mX{\mathbb{X}}

\def\mY{\mathbb{Y}}
\def\mZ{\mathbb{Z}}

\def\argmin{\mbox{argmin}}

\def\bI{\mbox{\boldmath $I$}}

\def\zero{\mathbf{0}}

\newcommand{\bfm}[1]{\ensuremath{\mathbf{#1}}}

   \def\bA{\bfm A}  
   \def\bB{\bfm B}  
\def\bc{\bfm c}

\def\bg{\bfm g}     
     
   \def\bI{\bfm I}

   \def\bW{\bfm W}

\def\wh{\widehat}

\def\boxit#1{\vbox{\hrule\hbox{\vrule\kern6pt\vbox{\kern6pt#1\kern6pt}\kern6pt\vrule}\hrule}}

\newcommand{\bfsym}[1]{\ensuremath{\boldsymbol{#1}}}

 \def\bmu{\bfsym {\mu}}                 
 
 \def\btheta{\bfsym {\theta}}           
           
 \def\bsigma{\bfsym \sigma}             \def\bSigma{\bfsym \Sigma}

 \def\btau{\bfsym {\tau}}

\renewcommand{\baselinestretch}{1.55}

\numberwithin{equation}{section}

\begin{document}

\def\spacingset#1{\renewcommand{\baselinestretch}%
{#1}\small\normalsize} \spacingset{1}


\if0\blind
{
  \title{\bf Graphical Assistant Grouped Network Autoregression Model: a Bayesian Nonparametric Recourse}
  \author{Yimeng Ren$^1$,  Xuening Zhu$^2$, Guanyu Hu$^3$\thanks{
    Yimeng Ren and Xuening Zhu are supported by the National Natural Science
    Foundation of China (nos. 11901105, 71991472, U1811461), the Shanghai Sailing Program for Youth Science and
    Technology Excellence (19YF1402700), and funds from Fudan University.
    Xuening Zhu is the corresponding author.
    }\hspace{.2cm}\\
    $^1$School of Statistics, Renmin University of China\\
    $^2$School of Data Science, Fudan University\\
    $^3$Department of Statistics, University of Missouri Columbia}
  \maketitle
}

\fi

\if1\blind
{
  \bigskip
  \bigskip
  \bigskip
  \begin{center}
    {\LARGE\bf Graphical Assistant Grouped Network Autoregression Model: a Bayesian Nonparametric Recourse}
\end{center}
  \medskip
} \fi
\bigskip
\begin{abstract}
Vector autoregression model is ubiquitous in classical time series data analysis. With the rapid advance of social network sites, time series data over latent graph is becoming increasingly popular. In this paper, we develop a novel Bayesian grouped network autoregression model to simultaneously estimate group information (number of groups and group configurations) and group-wise parameters. Specifically, a graphically assisted Chinese restaurant process is incorporated under framework of the network autoregression
model to improve the statistical inference performance.  
An efficient Markov chain Monte Carlo sampling algorithm is used to sample from the posterior distribution. Extensive studies are conducted to evaluate the finite sample performance of our proposed methodology. Additionally, we analyze two real datasets as illustrations of the effectiveness of our approach.
\end{abstract}

\noindent%
{\it Keywords:}  Graphical Assistant Chinese Restaurant Process, MCMC, Mixture Models,  Network Data, Vector Autoregression.
\vfill

\newpage
\spacingset{1.45}

\noindent
\section{Introduction}\label{sec:intro}

The network data is becoming increasingly popular, which has many important 
applications in various disciplines,
including sociology \citep{simmel1950sociology,wasserman1994social,hanneman2005introduction}, genomics \citep{wu2010human,horvath2011weighted}, psychology \citep{borgatti2009network,cramer2010comorbidity,borsboom2013network}, finance and economics \citep{zou2017covariance,zhu2019network, leung2017network}.
In the field of sociology, the network data is used to study the structure of relationships by linking individuals or other social units and by modeling interdependencies in behaviors or attitudes related to configurations of social relations \citep{o2008analysis}. 
In biological and genetic areas, the presence of an interaction between two genes or proteins indicates a biologically functional relationship \citep{von2002comparative}, which makes the network data widely used in cancer and other disease data analysis with promising results \citep{guda2009comparative, chuang2007network}. 
A continuous response observed on each node at equally spaced time points is very common in various applications such as biological studies, economics, and finance. Building a network based model to recover the dynamics of the responses is necessary.

We consider a network with $N$ nodes, which are indexed as $i = 1,\cdots, N$.
To represent the network relationship among the network nodes,
we employ an adjacency matrix $\bA = (a_{ij})\in \mR^{N\times N}$, where 
$a_{ij} = 1$ indicates the $i$th node follows the $j$th node, otherwise $a_{ij} = 0$. Following the convention we let $a_{ii} = 0$ for $1\le i\le N$.
Denote $Y_{it}$ as the continuous response
collected from the node $i$ during $1\le t\le T$.
Correspondingly, we collect a number of nodal covariates (e.g., user's age, gender), which is denoted by $V_i\in \mR^p$.
To model the dynamics of the response $Y_{it}$, \cite{zhu2017network} proposed a network vector autoregression (NAR) model, which can be expressed as follows,
\begin{align}
    Y_{i t}=\beta_{0}+\beta_{1} n_{i}^{-1} \sum_{j} a_{i j} Y_{j(t-1)}+\beta_{2} Y_{i(t-1)}+V_{i}^{\top} \gamma+\varepsilon_{i t},\label{nar}
\end{align}
where $n_i = \sum_j a_{ij}$ is the out-degree of node $i$
and $\ve_{it}$ is the random noise term.
The NAR model as well as its extensions
have been implemented to a broad range of fields, including spatial data modeling \citep{lee2009spatial, shi2017spatial}, social studies \citep{sojourner2013identification,liu2017peer,zhu2018grouped}, financial risk management \citep{hardle2016tenet,zou2017covariance}, and many others.

Despite the usefulness of the NAR model, it lacks flexibility in its model formulation and suffers from model misspecification.
Specifically, it specifies a homogeneous network autoregression coefficient
(i.e., $\beta_1$) for all network nodes, 
which is highly restrictive in practice.
To enhance the model's flexibility, a popular way is to consider a 
group structure of model coefficients \citep{ke2015homogeneity,su2016identifying,ando2016panel,gudhmundsson2021detecting}.
With respect to the network data, 
\cite{zhu2018grouped} considered a grouped network autoregression (GNAR) model.
To be more specific, the nodes are clustered into several groups, and the nodes within the same group share a set of common parameters.
Both the group memberships and group-specific regression coefficients are to be estimated.


Although the model's flexibility is enhanced and the misspecification risk is reduced, 
there are several unsolved challenges for the GNAR model. 
First, it is necessary to incorporate the graphical information (i.e., $\bA$) into grouping process. Nevertheless, most existing methods ignore graphical information in their grouping process. 
On the other hand, another line of research takes advantages of the graphical information and uses nonparametric Bayesian methods for clustering nodes \citep{sewell2017latent,geng2019probabilistic,van2018bayesian}, but they totally 
ignore the dynamics of $\{Y_{it}\}$.
The graphical information, a connectivity pattern among different nodes, will partially determines the group configurations. 
In other words, the closely connected nodes should have higher probability belonging to the same group. 
How to incorporate the graphical information into the modelling procedure of the GNAR model
is a challenging task.
Second, proposing a simultaneously inference procedure on both group configurations and group-wised parameters is another important consideration for GNAR model.
The uncertainty of parameter estimation is often neglected in existing literature, and it is difficult to get the probabilistic interpretations from existing frequentist approaches. 
Lastly, another important problem in grouping methods is how to determine the number of groups. 
Most existing methods require specification of the number of groups first and then estimate group configurations. 
The number of groups is determined by information criteria or some {\it ad hoc} procedures. These procedures may ignore uncertainty of the estimation of the number of groups in the first stage and lack uncertainty quantification of the number of groups, which is also an important issue for GNAR model.

Our methodology development is directly motivated by solving the aforementioned challenges to fill the gap between nonparametric Bayesian methods and network autoregression model. 
In this article, we propose a novel graphical assistant grouped network autoregression (GAGNAR) model, considering both the heterogeneous network effects across nodes and the graph information in the grouping procedure under a novel nonparametric Bayesian mixture model framework. 
Specifically, a graphically assistant Chinese Restaurant Process (gaCRP) borrowed from \citet{geng2021bayesian} is introduced to the GNAR model framework, which incorporates the graphical information and utilizes the sampling procedure from the Chinese Restaurant Process (CRP) \citep{pitman1995exchangeable}.
The proposed prior will provide simultaneous inference on the number of groups and group configurations. 
Moreover, the gaCRP has a P\'{o}lya urn scheme similar to the traditional CRP. Due to this merit, we can develop a Gibbs sampler that enables efficient full Bayesian inference on the number of groups, mixture probabilities as well as group-specific parameters. We demonstrate its excellent numerical performance through simulations and analysis of two real datasets, compared with several benchmark methods.

Our proposed method is unique in the following aspects. First, the proposed method provides a useful model-based solution for GNAR model that is able to leverage graphical information without pre-specifying the number of groups. Secondly, by adopting a full Bayesian framework, the grouping results yield useful probabilistic interpretation. Thirdly, the developed posterior sampling scheme also renders efficient computation and convenient inference since our proposed collapsed Gibbs sampling algorithm by marginalizing over the number of groups avoid complicated reversible jump Markov Chain Monte Carlo (MCMC) algorithm \citep{green1995reversible} or allocation samplers.

The rest of this paper is organized as follows. In Section \ref{sec:method}, we briefly review the NAR and GNAR models, followed by a review of Bayesian mixture model and Dirichlet process. In addition, we describe the graphical assistant prior model and introduce our proposed GAGNAR model and its theoretical properties. 
The Bayesian inferences including the MCMC sampling algorithm, the post-MCMC estimation and the model selection criterion for tuning parameter are presented in Section~\ref{sec:bayesian_inf}.
In Section \ref{sec:simu}, we study the finite sample performance of our proposed method via extensive simulation studies. Two real datasets are illustrated by our proposed method in Section \ref{sec:real_dat}. Section \ref{sec:discussion} concludes and raises some in-depth discussions. All technique proofs and additional numerical results are given in supplementary materials.


\section{Methodology}\label{sec:method}
\subsection{Grouped Network Autoregression Model}

Consider a network with $N$ nodes
with the corresponding adjacency matrix denoted by $\bA$.
Assume the nodes are clustered into $K$ latent groups.
Denote $\mathcal{F}_t$ as the $\sigma$-field generated by $\{ Y_{it}: 1 \le i \le N, 1 \le t \le T \}$. Given $\mathcal{F}_{t-1}$, the observations at time point $t$ are assumed to be independent and follow a Gaussian mixture distribution with $K$ components. Given the number of groups $K$
and the adjacency matrix $\bA$, the GNAR model can be expressed as
a mixture model, i.e., 
\begin{equation}
  \begin{split}
          &p(Y_{it}|\btheta, \sigma^2) = \sum_k \pi_k \mathcal{N}(\mu_{kt}, \sigma_k^2), \\
    &\mathcal{N}(\mu_{kt}, \sigma_k^2) = \frac{1}{\sqrt{2 \pi}\sigma_k} \exp\Big\{-\frac{(Y_{it}-\mu_{kt})^2}{2 \sigma_k^2}\Big\},\\
    &\mu_{kt} = \beta_{0k} +\beta_{1 k} n_{i}^{-1} \sum_{j} a_{i j} Y_{j(t-1)}+\beta_{2 k} Y_{i(t-1)}+V_{i}^{\top} \gamma_{k},\label{gnar-model}
  \end{split}  
\end{equation}
where $\mathcal{N}(\mu_{kt}, \sigma_k^2)$ is the density function of normal distribution with mean $\mu_{kt}$ and variance $\sigma_k^2$ in the $k$th group, and $\pi_k$ is the mixing weight satisfying $\sum_k \pi_k = 1$.
Here 
$\btheta_k =  (\beta_{0 k}, \beta_{1 k}, \beta_{2 k}, \gamma_{k}^{\top})^\top$ collects the unknown parameters.


The GNAR model \eqref{gnar-model} characterizes the conditional mean of the response $Y_{it}$ by a combination of  four components. 
The first one is the \textit{baseline effect} $\beta_{0k}$, which is a constant but
varying among different groups.
The second one is the network component ($\beta_{1k}n_i^{-1}\sum_j a_{ij}Y_{j(t-1)}$), which reflects the average impact of following nodes at the previous time point. 
Therefore, we refer to $\beta_{1k}$ as the \textit{group network effect}.
The third one is the momentum component ($\beta_{2k}Y_{i(t-1)}$). It characterizes how the focal node is influenced by its historical behaviors. The corresponding parameter $\beta_{2k}$ is then referred to as {\it group momentum effect}, quantifying the node's self-dependence.
The last one is the nodal covariate component ($V_i^\top\gamma_k$), which includes the node specific characteristics
and it is invariant over time.
As implied by the GNAR model (\ref{gnar-model}), the nodes within the same group share the same set of coefficients, therefore they exhibit 
similar dynamic patterns.
Compared to the NAR model \citep{zhu2017network}, the GNAR model is able to capture the nodes' heterogeneous pattern in the group level.
To estimate the model parameters simultaneously with the group memberships, an EM algorithm could be designed \citep{zhu2018grouped}.


Despite the usefulness, the model has three main limitations.
First, the estimation can be unreliable when the observed time length is short.
That decreases the model's stability and robustness in practice.
Second, the network structure information is not fully exploited for estimating the group memberships.
Third, the number of groups $K$ cannot be automatically estimated but needs to be specified in advance. 
That increases the possibility that the model is misspecified.
As a consequence, we are motivated to consider the problem in a Bayesian framework and introduce a graph assisted prior, 
where the network structure information is employed when forming the groups.
The details are presented in the following section.

\subsection{Bayesian Graph Assisted Prior}\label{subsec:prior}

For each node $i$, suppose it carries a latent group membership $z_{i}\in \{1,\cdots, K\}$, where $z_{i} = k$ implies that 
the node $i$ is from the $k$th group.
Given $\{z_{i}\}$ and  $\{\btheta_{k}, \sigma_k^2:1\le k\le  K\}$,
we have a Gaussian mixture model \citep{marin2005bayesian} as follows,
\begin{align*}
    &(Y_{it}|z_i, \btheta_{z_i}, \sigma_{z_i}^2) \sim \mathcal{N}(\mu_{z_it}, \sigma_{z_i}^2),\quad i=1,\ldots,N,\quad t=1,\ldots,T,
\end{align*}
where $\mu_{z_it}= \beta_{0z_i} +\beta_{1 z_i} n_{i}^{-1} \sum_{j} a_{i j} Y_{j(t-1)}+\beta_{2 z_i} Y_{i(t-1)}+V_{i}^{\top}\gamma_{z_i}$. Under the Bayesian hierarchical model, 
for the unknown group memberships, we specify the model priors as
\begin{equation}
\label{eq:gmm}
    \begin{split}
        &z_i\sim\text{Cat}(\pi_1,\ldots,\pi_K),\quad i=1,\ldots,N,\\
        &\pi_1,\ldots,\pi_K\sim \text{Dirichlet}_K(\alpha,\ldots,\alpha),\\
        &\btheta_j \sim \pi(\btheta), \quad j=1,\ldots,K,
    \end{split}
\end{equation}
where $z_i$ follows a categorical distribution with parameters $\pi = \{\pi_1,\dots \pi_K\}$, $\pi$ follows a Dirichlet distribution with parameter $\alpha$, and $\pi(\btheta)$ is joint prior for all group-wise parameters. When $K$ is unknown, we can let $K$ go to infinity, and then the model in \eqref{eq:gmm} becomes the Dirichlet process mixture model \citep[DPMM;][]{ishwaran2002exact}. 
The DPMM is very flexible and efficient for density estimation.
However, it suffers from several problems in estimating the group structure.
First, the partitions sampled from the DPMM posterior tend to have multiple small transient groups, which decreases the interpretability of the model.
Second, the DPMM is inconsistent on the estimation of the number of groups \citep{miller2018mixture}. 
Third, the DPMM does not take advantage of the network structure information: the connected nodes do not have higher probability being in the same group.

In order to address the above problems,
we employ a Bayesian graph assisted prior.
Specifically
we assume that 
the group membership $z_i \in \{1,\cdots, K\}$ is generated from the weighted Dirichlet process.
The process assumes that the nodes' memberships are sequentially generated. 
Suppose the nodes $1\le j\le i-1$ form $K'$ groups, then the 
conditional distribution for $z_i$ is expressed as
\begin{align}
P(z_{i}=k \mid z_{1}, \ldots, z_{i-1}) \propto\Bigg\{\begin{array}{ll}
p_k(\bW), & 1\le k\le K^{\prime}\\
\alpha, &  k = K^{\prime}+1
\end{array}\label{p_k}
\end{align}
where $p_k(\bW) = \sum_{j = 1}^{i-1} w_{ij} I(z_j = k)$
and $\bW = (w_{ij})$ is a weighting matrix calculated based on $\bA$.
The $w_{ij}$ in the weight matrix denotes the connection strength between node $i$ and $j$.
The parameter $\alpha$ controls the probability for introducing a new group.
As a result, \eqref{p_k} implies that a node is more likely to fit in the same group with its connected friends.

\begin{remark}\label{remark.1}
With the probabilistic Bayesian framework, simultaneous estimation of the number of groups and the grouping configuration is typically achieved by complicated searching algorithms (e.g., reversible jump MCMC \citep{green1995reversible}).
However, it suffers from high computational complexity.
    In our framework, the specification of \eqref{p_k} allows the nodes' memberships to be generated sequentially and therefore avoids setting the number of groups in advance.
    According to \eqref{p_k}, a new group will formulate once $z_i$ is assigned with a new group label $K'+1$.
    This enables a sequential generation of groups. Once we generate all the group memberships $\{z_i\}$, we know immediately the number of groups $K$.
\end{remark}
Let $G_0$ be a continuous probability measure on $\mathbb{R}^{p} \times \mathbb{R}^+$. We define the full conditional distribution of $\btheta_{z_i}$ given $\btheta_{z_1},\cdots,\btheta_{z_{i-1}}$ as 
\begin{align}
	f(\btheta_{z_i}\mid
	\btheta_{z_1},\ldots,\btheta_{z_{i-1}}) 
	 \propto \sum_{r=1}^{K^*}
\sum_{j=1}^{i-1}w_{ij} I (\btheta_{z_i}=\btheta^*_r) f_{\btheta_r^*}(\btheta_{z_i}) + \alpha G_0(\btheta_{z_i}),
\label{eq:conditional_distribution}
\end{align}
where 
$f(\cdot)$ is the density function,
$K^*$ denotes the number of groups excluding the $i$-th observation,
$\btheta^*_1,\ldots,\btheta^*_{K^*}$ are $K^*$ distinguished values of
$\btheta_{z_1},\ldots,\btheta_{z_{i-1}}$ and $I(\cdot)$ is the identity function. 
In practice we set the base distribution $G_0$ to be the normal inverse gamma distribution.

Inspired by the graph assisted Chinese restaurant process (gaCRP)  \citep{geng2021bayesian} for survival model, we define the graph distance $d_{ij}$ between node $i$
and $j$ as the shortest path length between node $i$ and $j$.
If node $i$ and $j$ cannot be connected with a finite length of path, then $d_{ij} = \infty$.
The weight is then defined as 
\begin{align}
w_{i j}=\Bigg\{\begin{array}{ll}
1, & \text { if } d_{ij} \leq 1 \\
\exp (-d_{ij} \times h), & \text { if } d_{ij}>1
\end{array}\Bigg. .\label{weight}
\end{align}
Therefore the weight $w_{ij}$ characterizes the closeness of the node $i$ and $j$ in the network.
The parameter $h$ denotes the smoothing scale of the weighting matrix.
A larger value of $h$ indicates the weight is more concentrating on the local scale.
For simplicity, we refer to gaCRP introduced above as gaCRP($\alpha, h$). Although the idea of using a weighted version of CRP has been explored in \cite{geng2021bayesian}, their model is based on Cox regression for survival data analysis, which significantly differs from our model, since it lacks discussion on dynamic process of $Y_{it}$.

\begin{remark}\label{remark.2}
    Implied by \eqref{weight}, if $h=0$, then the gaCRP is the same as the traditional CRP and may tend to over cluster the nodes. If $h \rightarrow \infty$, the probability for a new node to choose the existing groups tends to be small, which may also lead to the over-clustering problem.
    Hence, appropriate selection of $h$ is critical for the model performance.
    We discuss in details about the selection of $h$ in Section \ref{subsec:select_model}.
\end{remark}

In summary, we express the proposed model hierarchically as 
\begin{equation}
    \begin{split}
        & \text{Data model}: Y_{it}|z_i, \btheta_{z_i}, \sigma^2_{z_i} \sim \mathcal{N}(\mu_{z_it}, \sigma_{z_i}^2),\\
    & \text{Process model}: p(z_i|z_{1}, \cdots, z_{i-1})\sim \text{gaCRP}(\alpha, h),\\
    & \text{Parameter model}: (\btheta_{z_i}, \sigma_{z_i}^2) \sim \text{NIG} (\btau_0, \bSigma_0, a_0, b_0), \label{eq:hierearchical_model}
    \end{split}
\end{equation}
where $\btau_0$, $\bSigma_0$, $a_0$ and $b_0$ are hyper-parameters for base distribution of $(\btheta_k, \sigma_k^2)$. 
Here the prior distribution of $(\btheta_{z_i}, \sigma_{z_i}^2)$ is set to be normal inverse gamma (NIG) distribution with parameters $(\btau_0, \bSigma_0, a_0, b_0)$.
Without loss of generality, we choose $\btau_0 = \zero$, $\bSigma_0 = 100\bI$, $a_0 = 0.01$, $b_0 = 0.01$ in our simulation study and real data applications. 
We refer to the model \eqref{eq:hierearchical_model} as graph assisted group network autoregression (GAGNAR) model.







\section{Bayesian Estimation}\label{sec:bayesian_inf}

In this section, we discuss the model estimation for the GAGNAR model. 
Define group membership vector $\mZ = (z_1,\cdots, z_N)^\top\in \mR^N$.
If the group memberships are known,
then the model parameters in the GNAR model (\ref{gnar-model}) can be easily estimated with least square
estimation method.
However, in practice $\mZ$ is unobserved thus the least square estimation is infeasible.
To estimate the model parameters simultaneously with the latent group memberships, we employ the Markov chain Monte Carlo (MCMC) algorithm with details given as follows.

\subsection{MCMC Algorithm}

To conduct Bayesian estimation, we first derive the 
posterior distribution of the unknown parameters and group memberships.
Define $ \btheta = (\btheta_1, \cdots, \btheta_K)$ and $ \bsigma^2 = (\sigma_1^2, \cdots, \sigma_K^2)$. 
In addition, denote $\mV = (V_1,\cdots, V_N)^\top\in \mR^{N\times p}$, 
$\mY_t = (Y_{1t},\cdots, Y_{Nt})^\top\in\mR^N$,
and $\mY = ( \mY_1,\cdots, \mY_T)\in \mR^{N\times T}$.
Given the data $\{\mY, \mV\}$, the joint posterior distribution of latent membership $\mZ$ and unknown parameters
$\{\btheta, \bsigma^2\}$ could be written as
\begin{align}
    \pi(\mZ, \btheta, \bsigma^2|\mY, \mV)\propto \pi(\mZ)\pi(\btheta)\pi(\bsigma^2)f(\mY|\mZ, \mV, \btheta,\bsigma^2),\label{posterior}
\end{align}
where $\propto$ denotes ``proportional to", and $\pi(\mZ)$, $\pi(\btheta)$, $\pi(\bsigma^2)$ stand for the prior distributions for $\mZ$,
$\btheta$, and $\bsigma^2$ respectively. 
For convenience, let $ \mathbb{X}_{t} = {(X_{1t},\cdots, X_{Nt})^\top} = (\one_N, \bW \mathbb{Y}_{t-1}, \mathbb{Y}_{t-1}, \one_N V_i^\top) \in \mR^{N\times (p+3)}$, then given $\btheta_k, \sigma_k^2$ we have 
\begin{align*}
    (\mathbb{Y}_{t} | \btheta_k, \sigma_k^2) \sim \mathcal{N}(\bmu_k, \sigma_k^2 \bI), \text{ where } \bmu_k = \mathbb{X}_{t} \btheta_k.
\end{align*}
The prior distribution for $\mZ$ is specified as gaCRP$(\alpha, h)$ as stated in Section \ref{subsec:prior}.
In addition, we specify the prior distribution for $\btheta_k$ as multivariate normal distribution $\mN(\btau_0, \sigma_k^2 \bSigma_0)$ and we set the prior for $\sigma_k^2$ as the commonly used scaled inverse gamma distribution $\text{IG}(a_0, b_0)$, where $\btau_0$, $\bSigma_0$, $a_0$ and $b_0$ are the hyper parameters.

Given the joint posterior distribution of $\{\mZ, \btheta, \bsigma^2\}$ in (\ref{posterior}), we can obtain the conditional posterior distribution for each latent group membership and model parameters respectively.
This enables us to conduct the MCMC algorithm for model estimation.
We first give the conditional posterior distribution for $z_i$ $(i = 1,\cdots, N)$ as follows.
For convenience we let $\mZ_{(-i)} = \{z_j: j\ne i\}$.
Denote full conditional of $Z_i$ as $\wt f(z_i) = f(z_i|\mZ_{(-i)}, \mY, \mV, \btheta, \bsigma^2)$.
Further define $\mY_{(i)} = (Y_{i1},\cdots, Y_{iT})^\top\in \mR^T$ and $\mX_{(i)} = (X_{i1},\cdots, X_{iT})^\top\in \mR^{T \times (p+3)}$.
Then we have
\begin{align*}
   \wt f(z_{i}) &\propto f(\mZ,\btheta, \bsigma^2|\mY,\mV) \propto f(\mY_{(i)}|z_i, \mV, \btheta,\bsigma^2) f(z_i|\mZ_{(-i)}).
\end{align*}
The analytical form of $\wt f(z_i)$ is given in Proposition \ref{prop.1}.
\begin{proposition}\label{prop.1}
Suppose currently the nodes are formed into $K'$ groups.
    Then  according to (\ref{eq:hierearchical_model}) we have
    \begin{align*}
\wt f(z_i) \propto\Bigg\{\begin{array}{ll}
\kappa_k f(\mY_{(i)};\mX_{(i)},\btheta_k, \sigma_k^2), & \text { for } 1\le k\le K' \\
\alpha g(\mY_{(i)};\btau_0, \btau, \bSigma_0, \bSigma, a_0,b_0), & \text { for } k = K'+1
\end{array} ,
\end{align*}
    where $\kappa_k = \sum_{j \neq i} w_{i j} I(z_j = k)$ and
    \begin{align*}
    &f(\mY_{(i)};\mX_{(i)},\btheta_k, \sigma_k^2) = (2 \pi \sigma_k^2)^{-(T-1)/2}\exp\Big\{-\frac{1}{2\sigma_k^2} \|\mY_{(i)} - \mX_{(i)} \btheta_k\|^2\Big\},\\
    &g(\mY_{(i)};\btau_0, \btau, \bSigma_0, \bSigma, a_0,b_0) = \varphi(\bSigma_0, \bSigma, a_0,b_0)
    \big\{ b_0+\frac{1}{2} ( \btau_0^\top \bSigma_0^{-1} \btau_0 + \mY_{(i)}^\top \mY_{(i)} - \btau^{\top} {\bSigma}^{-1} \btau ) \big\} ^{-(a_0 + \frac{T-1}{2})} ,\\
    & \varphi(\bSigma_0, \bSigma, a_0,b_0) = \frac{b_0^{a_0} \Gamma(a_0 + \frac{T-1}{2}) |{\bSigma}|^{1/2}}{(2 \pi)^{(T-1)/2} \Gamma(a_0) |\bSigma_0|^{1/2}} ,\\
    &{\bSigma} = (\bSigma_0^{-1} + \mX_{(i)}^\top \mX_{(i)})^{-1},~~~ \btau = \bSigma (\bSigma_0^{-1}\btau_0 + \mX_{(i)}^\top \mY_{(i)}).
    \end{align*}
\end{proposition}

By the formulation of $\wt f(z_i)$,
we can observe that the probability $z_i$ belonging to an existing group $k$ ($1\le k \le K'$) is closely related to the network weighting matrix $\bW$.
Particularly $\kappa_k$ denotes a weighted average ratio
of its connected friends belonging to the group $k$,
which characterizes a {\it stickiness} to the group $k$ of $i$th neighbourhood.
If the stickiness level is higher, then the conditional probability of $z_i = k$ will be higher.
That is how the network topology information is involved in the sampling procedure.
Next, the node is allowed to fit into a new group with probability proportional to $g(\mY_{(i)};\btau_0, \btau, \bSigma_0, \bSigma, a_0,b_0)$,
which relates to both prior information and 
data information.

Next, we investigate the full conditional distribution for model parameters $\{\btheta, \bsigma^2\}$. Denote the posterior distribution of $(\btheta_k, \sigma_k^2)$ as $\wt{f}(\btheta_k, \sigma_k^2) = f(\btheta_k, \sigma_k^2 | \mZ, \mY, \mV)$, then we derive $\wt{f}(\btheta_k, \sigma_k^2)$ in the following Proposition \ref{prop.2}.

\begin{proposition}\label{prop.2}
    The full conditional distribution $\wt{f}(\btheta_k, \sigma_k^2) (k=1,2,\cdots,K)$ is given as 
    \begin{align*}
    \wt{f}(\btheta_k, \sigma_k^2) & \propto \pi(\btheta_k, \sigma_k^2 ; \btau_0, \bSigma_0, a_0, b_0) \prod_{z_i = k} f(\mY_{(i)} ;\mX_{(i)},\btheta_k, \sigma_k^2) \\
    & \propto f(\btheta_k, \sigma_k^2 ; \btau^*, \bSigma^*, a^*, b^*),
    \end{align*}
    where 
    \begin{align*}
    & f(\btheta_k, \sigma_k^2 ; \btau_0, \bSigma_0, a_0, b_0)\\
    &= \frac{b_0^{a_0}}{(2 \pi)^{d/2} | \bSigma_0 |^{1/2} \Gamma(a_0)} \Big(\frac{1}{\sigma_k^2}\Big) ^{a_0 + \frac{d}{2} + 1} \exp \Big[ -\frac{1}{\sigma_k^2 } \Big\{ b_0 + \frac{1}{2} (\btheta_k - \btau_0)^\top \bSigma_0^{-1} (\btheta_k - \btau_0) \Big\} \Big] ,\\
    & \btau^* = (\bSigma_0^{-1} + \sum_{z_i = k} \mX_{(i)}^\top \mX_{(i)})^{-1} (\bSigma_0^{-1}\btau_0 + \sum_{ z_i = k} \mX_{(i)}^\top \mY_{(i)}), ~~~ \bSigma^* = (\bSigma_0^{-1} + \sum_{ z_i = k} \mX_{(i)}^\top \mX_{(i)})^{-1}, \\
    & a^* = a_0 + (T-1) N_k /2, ~~~ b^* = b_0 + \frac{1}{2}\Big[\btau_0^\top \bSigma_0^{-1} \btau_0 + \sum_{z_i = k}\mY_{(i)}^\top \mY_{(i)} - \btau^{*\top} \bSigma^{* {-1}} \btau^* \Big], \\
    & \text{and} \ N_k = \sum_i I(z_i = k).
    \end{align*}
\end{proposition} 
The proofs of Propositions are given in supplementary materials. 
As shown in Proposition \ref{prop.2}, the conditional distribution of $\{\btheta_k,\sigma_k^2\}$
follows normal inverse gamma distribution (NIG) with 
density function 
$f(\btau^*, \bSigma^*, a^*, b^*)$.
With the expressions of $(\btau^*, \bSigma^*, a^*, b^*)$, we see that it unifies the prior information $\{\btau_0,\bSigma_0, a_0,b_0\}$ and the data information from the $k$th group, i.e.,  
$\{\mX_{(i)}, \mY_{(i)}: z_i = k\}$.
Based on Proposition \ref{prop.1} and Proposition \ref{prop.2}, we could efficiently cycle through the full conditional distribution $\wt f(z_i)$ for $i = 1,2,\cdots, N$ and $\wt{f}(\btheta_k, \sigma_k^2)$ for $k=1,2,\cdots,K$ to conduct Gibbs sampling procedure.
One should note that in the process of node memberships generation,
$K$ is marginalized over, which could avoid complicated reversible jump MCMC algorithms or allocation samplers.
We summarize the collapsed Gibbs sampling procedure for the GAGNAR model in Algorithm \ref{alg1}.

\begin{algorithm}
\caption{Collapsed Gibbs sampler for GAGNAR}
\begin{algorithmic}[1]
\State Observations $\mY_{(1)}, \mY_{(2)}, \dots, \mY_{(N)}$.
\State Initialize the number of groups $k$, the group assignment $z_i$ and the parameters $(\btheta_{z_i}, \sigma_{z_i}^2)$ for each $\mY_{(i)}$
\For {$iter $ from $ 1$ to $M$}
    \For {$i$ from $1$ to $N$}
        \State Remove $\mY_{(i)}$ from group $z_i$. If $\mY_{(i)}$ is from a new cluster, set $k = k-1$.
        \State Update $(z_1, \dots, z_N)$ conditional on $\btheta, \sigma^2$ for each node $i \in (1,\dots,N)$ by the conditional distribution $\wt f(z_i)$ in Proposition \ref{prop.1}.
    \EndFor
    \For {$c$ from $1$ to $K$}
    \State Update $(\btheta_c, \sigma_c^2)$ for each $c \in (1,\dots, K)$ conditional on $(\mZ, \mY, \mV)$ by the density function in Proposition \ref{prop.2} as
    \begin{align*}
        \wt{f}(\btheta_c, \sigma_c^2) \sim \text{NIG}(\btau^*, \bSigma^*, a^*, b^*)
    \end{align*}
    \EndFor
\EndFor
\end{algorithmic}\label{alg1}
\end{algorithm}

\subsection{Post MCMC Estimation}\label{subsec:inference}

Let $\btheta_k^{(m)}$ and $\sigma_k^{2(m)}$ with $1\le k\le K^{(m)}$ be the $m$th estimate after the burn-in iterations of the MCMC algorithm.
Correspondingly denote $\mZ^{(m)} = (z_i^{(m)}: 1\le i\le N)^\top$ as the estimated memberships of the network nodes.
Particularly we note that the estimated number of groups $K^{(m)}$ can be different for $1\le m\le M$.
As a result, direct posterior estimation of the parameters would be difficult since the number of estimated parameters are not the same across different iterations.
To address this issue, we adopt Dahl's method \citep{dahl2006model} to select the best post burn-in iteration with the least squares criterion.
The estimate output by the best post burn-in iteration is then taken as the final post MCMC estimator.


Specifically, we take advantage of the co-membership matrix to help us select the best post burn-in iteration.
Define the co-membership matrix as $\bB = (b_{ij})\in \mR^{N\times N}$,
where $b_{ij} = I(z_i = z_j)$.
Therefore the $(i,j)$th element of $\bB$ denotes whether the $i$th node is in the same group with the $j$th node.
We can see that $\bB$ is well defined for different $K^{(m)}$s and it does not suffer from the label switching issue.
We then choose the best post burn-in iteration as
\beq
m_b = \argmin_{1\le m\le M} \big\|\bB^{(m)}-\ol{\bB}\big\|_F^{2},\nonumber
\eeq
where $\bB^{(m)}$ is the estimated co-membership matrix in the $m$th post burn-in iteration and
$\ol \bB = M^{-1}\sum_{m = 1}^M \bB^{(m)}$.
As a consequence, the best post burn-in iteration is selected by the closest $\bB^{(m)}$ to the mean grouping co-membership $\ol\bB$.
The number of groups is then determined as $K^{(m_b)}$.
Accordingly, the post MCMC estimations of the parameters and memberships are given by 
$(\btheta_k^{(m_b)}, \sigma_k^{2(m_b)})$ with $1\le k\le K^{(m_b)}$ 
and $\mZ^{(m_b)}$ respectively.

\subsection{Selection of Smoothing Parameter $h$}\label{subsec:select_model}

The selection of smoothing parameter $h$ is redesigned as a model selection problem. 
Specifically, we use the logarithm of the pseudo marginal likelihood (LPML) \citep{ibrahim2014b} based on
conditional predictive ordinate (CPO) \citep{gelfand1992model, geisser1993predictive, gelfand1994bayesian} to select $h$.

The LPML given a specified $h$ is defined as 
\begin{align}
\mathrm{LPML}(h)=\sum_{i=1}^{N} \log \{\mathrm{CPO}_{i}(h)\}\label{lpml},
\end{align}
where CPO$_i(h)$ 
is the CPO for the node $i$ defined as 
CPO$_{i}(h) = p(\mY_{(i)} | \mY_{(-i)})$ \citep{pettit1990conditional}, 
and $\mY_{(-i)} = \{\mY_{(j)}: j \ne i\}$.
As one can see, LPML is a pseudo log-likelihood function and we prefer 
a smoothing parameter $h$ to maximize
LPML.
Borrowing the idea of \cite{chen2012monte}, we obtain the Monte Carlo estimate of CPO within the Bayesian framework as
\begin{align*}
\widehat{\mathrm{CPO}}_{i}(h)
=\Big\{\frac{1}{M} \sum_{m=1}^{M} \frac{1}{L(\btheta^{(m)}_{z_{i}}, \sigma^{2(m)}_{z_{i}};h)}\Big\}^{-1},
\end{align*}
where $M$ is the total number of Monte Carlo iterations, and  
\begin{align*}
    L(\btheta_{z_i}^{(m)}, \sigma_{z_i}^{2(m)}; h) &= \prod_{t=2}^T p(Y_{it} | \btheta_{z_i}^{(m)}, \sigma_{z_i}^{2 (m)};h )
\end{align*}
is the likelihood function for the node $i$
with the specified smoothing parameter $h$.
Correspondingly we have 
$\wh{\mathrm{LPML}}(h)=\sum_{i=1}^{N} \log \{\wh{\mathrm{CPO}}_{i}(h)\}$
and we choose the optimal $h$ by
$h_b = \arg\max_h \wh{\mathrm{LPML}}(h)$.



\section{Simulation}\label{sec:simu}
\subsection{Simulation Models and Data}

To demonstrate the finite sample performance of our proposed method, we conduct a number of numerical studies in this section. Specifically, we utilize three types of graph and assign two parameter settings to each of them.

For each simulation setting, we generate the random noise $\ve_{it}$ from a standard normal distribution. In addition, node covariates $V_i \in \mathbb{R}^{p}$ are independently sampled from a multivariate normal distribution $\mN(\zero, \bI_p)$. For each graph, two different scenarios of  parameters $(\beta_{0k}, \beta_{1k}, \beta_{2k}, \gamma_k, \sigma_k^2)$, which are listed in Table \ref{table:simu_para}. 
Given the initial value $\mathbb{Y}_{0}=\mathbf{0}$, the time series $\mathbb{Y}_{t}$ is generated according to the grouped network vector auto-regression model in \eqref{gnar-model}. 
In each scenario, 100 replicated datasets are generated,
and in each replicate we set the time length $T=20$. 
In addition, for the prior distribution gaCRP$(\alpha, h)$, we set $\alpha =1$ and select the smoothing parameters $h \in \{0,0.2,0.4,\cdots,5.0\}$ by LPML \eqref{lpml}.
A total of 1500 MCMC iterations are run for each replicate, with the first 500 iterations treated as burn-in. 
Besides, the hyper-parameters set for the base distribution in \eqref{eq:hierearchical_model} are 
$(a_0,b_0, \btau_0, \bSigma_0) = (0.01, 0.01, \zero, 100 \bI)$, which are noninformative priors.

\begin{table}[h!]
\centering
\caption{Simulation parameters setting for three examples}
\scalebox{0.8}{
\begin{tabular}{l|ccccc|ccccc} 
\toprule
& $\sigma^2$ & $\beta_{0}$ & $\beta_{1}$ & $\beta_{2}$ & $\gamma$ &$\sigma^2$ & $\beta_{0}$ & $\beta_{1}$ & $\beta_{2}$ & $\gamma$ \\
\midrule
\multicolumn{11}{c} {\textbf{Example 1}} \\
 \midrule
 & \multicolumn{5}{c} { Scenario 1 } &\multicolumn{5}{c} { Scenario 2 } \\
\midrule
Group 1 &2.0& $5.0$ & $0.2$ & $0.1$ & $(0.5,0.7,1.0)^{\top}$&2.0& $0.0$ & $0.1$ & $0.3$ & $(0.5,0.7,1.0)^{\top}$\\
Group 2 &1.0& $-5.0$ & $-0.4$ & $0.2$ & $(0.1,0.9,0.4)^{\top}$&4.0& $0.2$ & $-0.3$ & $0.2$ & $(0.1,0.9,0.4)^{\top}$\\
Group 3 &3.0& $0.0$ & $0.2$ & $0.4$ & $(0.2,-1.0,2.0)^{\top}$&3.0& $0.5$ & $0.2$ & $0.7$ & $(0.2,-0.2,1.4)^{\top}$\\
\midrule
\multicolumn{11}{c} {\textbf{Example 2 }} \\
\midrule
& \multicolumn{5}{c} { Scenario 1 } &\multicolumn{5}{c} { Scenario 2 } \\
\midrule
Group 1 &2.0 & $5.0$ & $0.2$ & $0.1$ & $(0.5,0.7,1.0)^{\top}$ & 2.0 & $0.0$ & $0.1$ & $0.3$ & $(0.5,0.7,1.0)^{\top}$ \\
Group 2 &1.0& $-5.0$ & $-0.4$ & $0.2$ & $(0.1,0.9,0.4)^{\top}$ &1.0 & $0.2$ & $-0.3$ & $0.2$ & $(0.1,0.9,0.4)^{\top}$ \\
Group 3 &3.0&  $0.0$ & $0.2$ & $0.4$ & $(0.2,-1.0,2.0)^{\top}$ &3.0 & $0.5$ & $0.2$ & $0.7$ & $(0.2,-0.2,1.4)^{\top}$ \\
Group 4 &4.0& $-0.1$ & $0.1$ & $0.2$ & $(1.0,-1.0,1.5)^{\top}$ &4.0 & $-0.1$ & $0.1$ & $0.2$ & $(1.0,-1.0,1.5)^{\top}$ \\
Group 5 &2.0& $3.0$ & $0.5$ & $0.2$ & $(0.8,0.5,-2.0)^{\top}$ &2.0 & $0.8$ & $0.5$ & $0.2$ & $(0.8,0.5,-1.0)^{\top}$ \\
\midrule
\multicolumn{11}{c} {\textbf{Example 3 }} \\
\midrule
& \multicolumn{5}{c} { Scenario 1 } & \multicolumn{5}{c} { Scenario 2 }\\
\midrule
Group 1 &2.0 & $5.0$ & $0.2$ & $0.1$ & $(0.5,0.7,1.0)^{\top}$&2.0 & $0.0$ & $0.1$ & $0.3$ & $(0.5,0.7,1.0)^{\top}$ \\
Group 2 &1.0& $-5.0$ & $-0.4$ & $0.2$ & $(0.1,0.9,0.4)^{\top}$&1.0 & $3.0$ & $-0.3$ & $0.2$ & $(0.1,0.9,0.4)^{\top}$ \\
Group 3 &3.0&  $0.0$ & $0.2$ & $0.4$ & $(0.2,-1.0,2.0)^{\top}$&3.0 & $-3.0$ & $0.2$ & $0.7$ & $(0.2,-0.2,1.4)^{\top}$ \\
Group 4 &4.0& $3.0$ & $0.1$ & $0.2$ & $(1.0,-1.0,1.5)^{\top}$&1.5 & $4.5$ & $0.1$ & $0.2$ & $(1.0,-1.0,1.5)^{\top}$ \\
Group 5 &2.0& $-3.0$ & $0.5$ & $0.2$ & $(0.8,0.5,-2.0)^{\top}$&2.5 & $-2.0$ & $0.5$ & $0.2$ & $(0.8,0.5,-1.0)^{\top}$ \\
Group 6 &3.0& $2.0$ & $-0.6$ & $-0.2$ & $(-0.8,0.5,2.0)^{\top}$&1.0 & $2.0$ & $-0.6$ & $-0.2$ & $(-0.8,0.5,2.0)^{\top}$ \\
\bottomrule
\end{tabular} 
}
\label{table:simu_para}
\end{table}

\textbf{Example 1.} (Stochastic Block Model) We first consider the stochastic block model (SBM) \citep{wang1987stochastic, nowicki2001estimation, zhao2012consistency}. 
The SBM assumes that nodes in the same group (block) are more likely to be connected, when compared with nodes from different groups. 
The model is widely used to discover community structures for network data.
We follow \cite{nowicki2001estimation} to randomly assign each node a group label $k=1,\cdots, K$ with equal probability $1/K$ and we set $K = 3$.
Next, let $P(a_{i j}=1)=20 N^{-1} $ if node $i$ and node $j$ are in the same group, and $P(a_{i j}=1)=2 N^{-1}$ otherwise.
For this example we set the network size $N = 100$.

\textbf{Example 2.} (Chinese Cities Graph) 
In this example
we construct the graph of Chinese cities by using the geographical information. 
We collect 151 cities of mainland China and treat each city as a network node in the graph.
The edge between two cities is defined as whether they share the common boarder
\citep{cao2017china, zhang2018technological}. 
Specifically,  $a_{ij} = 1$ illustrates that two cities are connected, otherwise $a_{ij} = 0$.
Lastly, we let $K=5$ be the total number of groups
and assign the group labels using $k$-means clustering on the adjacency matrix.

\begin{figure}[htpb!]
\centering
    \includegraphics[width=0.3\textwidth]{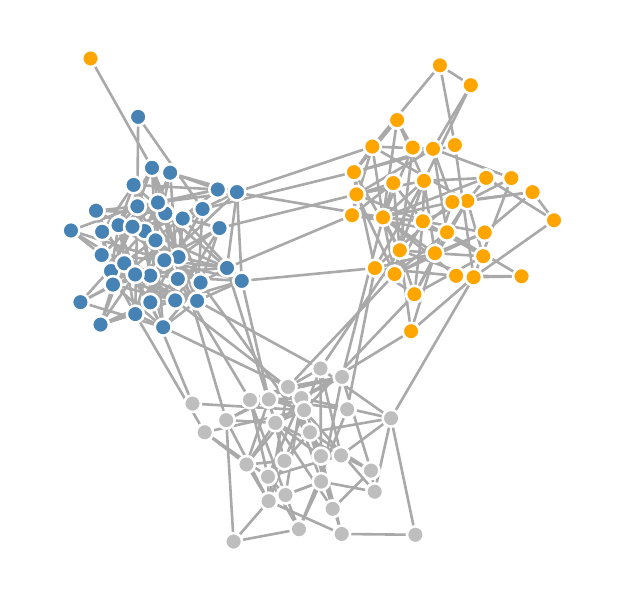}
    \includegraphics[width=0.3\textwidth]{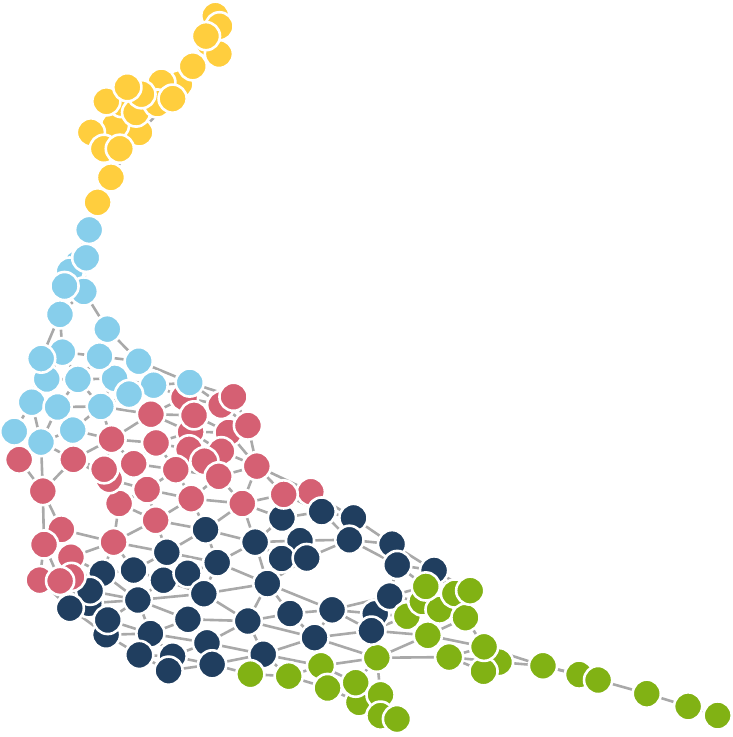}
    \includegraphics[width=0.3\textwidth]{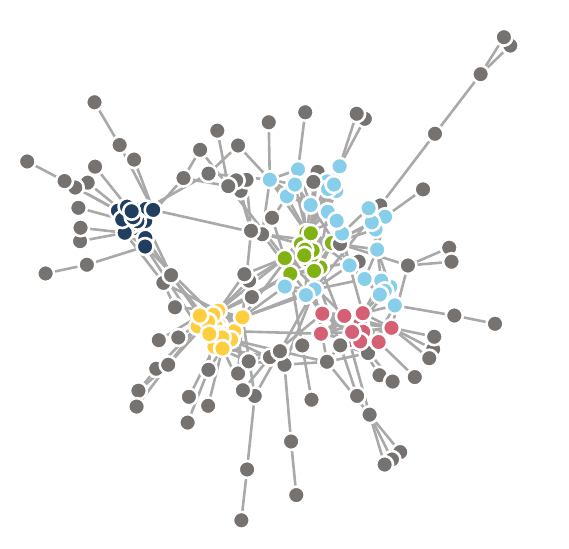}
\caption{\small Three graphs with true group labels, each color representing a group. Left: graph generated from SBM with $K=3$. Middle: geographical graph of $N = 151$ Chinese cities with $K=5$. Right: common shareholder graph of $N = 180$ stocks with $K=6$.}
\label{fig:sim_g}
\end{figure}

\textbf{Example 3.} (Common Shareholder Network) We lastly construct the stock graph from Chinese A stock market, which contains $N = 180$ actively traded stocks in the 
Shanghai and the Shenzhen Stock Exchange. 
The stocks used in this simulation example is a subset of our empirical study for computational convenience.
Specifically, $a_{ij} = 1$ indicates that two stocks share at least one of the top two shareholders, otherwise $a_{ij} = 0$.
The graph structure is shown in the bottom right panel of 
Figure \ref{fig:sim_g}. 
We set the number of groups $K = 6$ and assign the group memberships by implementing the SBM to fit the adjacency matrix.

The three network structures with the true group labels are visualized in Figure \ref{fig:sim_g}.


\subsection{Performance Measurements and Simulation Results}

Let $(\hat{\beta}_{0 k}^{(r)}, \hat{\beta}_{1 k}^{(r)}, \hat{\beta}_{2 k}^{(r)}, \hat{\gamma}_{k}^{(r)}, \hat{\sigma}_k^{2(r)})$ be the estimated parameters of the $k$th group by the $r$th replicate ($1\le r\le R$).
For each node $i$, we obtain its group label as $\hat{z}_{i}^{(r)} (i = 1, \cdots, N)$ by Dalh's method. 
Subsequently, the estimated parameters for each node $i$ is given as  $\big(\hat{\beta}_{0\wh z_i^{(r)}}^{(r)}, \hat{\beta}_{1\wh {z}_i^{(r)}}^{(r)}, \hat{\beta}_{2\wh {z}_i^{(r)}}^{(r)}, \hat{\gamma}_{\wh {z}_i^{(r)}}^{(r)}, \hat{\sigma}_{\wh{z}_i}^{2(r)}\big)$.
We consider the following measurements to evaluate 
the finite sample performance.
First, we employ the root mean square error (RMSE) to evaluate the estimation accuracy. For $\beta_{sk} $ ($0\le s\le 2, 1\le k\le K$), the RMSE is defined as  
\begin{align*}
   \operatorname{RMSE}_{\beta_{s}} = \Big\{(R N)^{-1} \sum_{r=1}^{R}\sum_{i=1}^{N} \big(\hat{\beta}_{s\wh z_i^{(r)}}^{(r)}-\beta_{sz_i}\big)^{2}\Big\}^{1 / 2},
\end{align*}
and the RMSE for $\gamma$ is calculated as 
\begin{align*}
  \operatorname{RMSE}_{\gamma}=\Big\{(R N)^{-1} \sum_{r=1}^{R} \sum_{i=1}^{N} \big\| \hat{\gamma}_{\wh{z}_i^{(r)}}^{(r)}-\gamma_{z_i} \big\|^{2}\Big\}^{1 / 2}.
\end{align*}
To measure the similarity between the estimated group memberships  and the true group memberships, we use Adjusted Rand Index (ARI) \citep{rand1971objective, hubert1985comparing}.
Define $\{z_i:1\le i\le N\}$ as the set of the true group memberships.
Then ARI is defined as 
\begin{align*}
&\text{ARI}^{(r)}=\frac{\text{RI}^{(r)}-E\big(\text{RI}^{(r)}\big)}{\max \big(\text{RI}^{(r)}\big)-E\big(\text{RI}^{(r)}\big)} , ~~~ \text{RI}^{(r)} = \frac{a^{(r)}+b^{(r)}}{C_N^2} ,
\end{align*}
where $a^{(r)} = \sum_{i,j} I(z_i = z_j, \wh z_i^{(r)} = \wh z_j^{(r)})$,
$b^{(r)} = \sum_{i,j} I(z_i \ne z_j, \wh z_i^{(r)} \ne \wh z_j^{(r)})$,
and $C_N^2 = N(N-1)/2$ is the total number of possible node pairs formed by the $N$ nodes.
Hence ARI measures the alignment level of two grouping results and a higher ARI value implies the estimated group memberships are more consistent with the true memberships.
The ARI can be calculated using the R-package \textit{mclust}.
We compare the parameter estimation of proposed model with that of EM algorithm and two-step estimation for the GNAR model \citep{zhu2018grouped}. 
One should note that, since the number of groups could not be inferred in the two baseline methods, we apply the true value of $K$ as the input of EM and two-step method.

Table \ref{table:SBM_simu_res1} summarises the average RMSE of estimated parameters under the optimal $h$ selected by LPML. 
For all three simulation examples, we see that the overall RMSEs of estimated parameters by the GAGNAR model are much lower than those of the EM and two-step methods.
This is consistent with the weakness we mentioned that the GNAR model does not utilize the network structure information.
In addition, we see that graphs in example 2 and 3 are more complicated than that generated from SBM, due to the larger number of groups and the group patterns which are not easily captured (see Figure \ref{fig:sim_g}). In both examples, the estimation accuracy of our model is shown to be much higher than that EM and two-step estimators. 
This further confirms the usefulness of the proposed GAGNAR model in the intricate reality.

\begin{figure}[htpb!]
    \centering
    \includegraphics[width=0.85\linewidth]{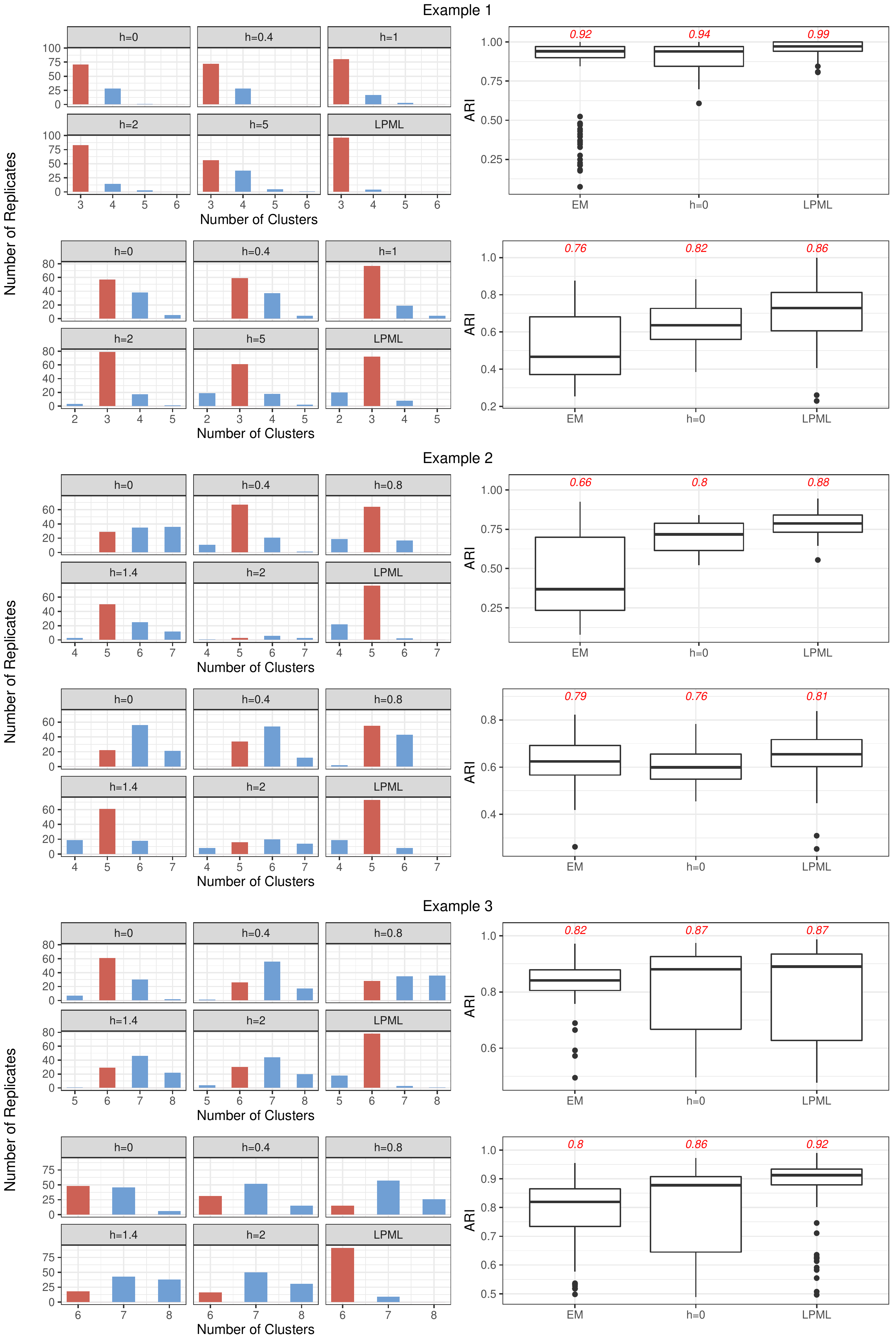}
    \caption{Grouping results of three examples, with results of scenario 1 in the first line and scenario 2 in the second line. Left panel: histogram of estimates of $K$ under different $h$. Right panel: boxplots of ARI under EM method, $h=0$ and optimal $h$. The red texts show the average accuracy of group memberships. The optimal $h$s selected by LPML are listed at the top of each example.}
    \label{fig:sim_res_merge}
\end{figure}

Figure \ref{fig:sim_res_merge} shows the estimated number of groups in the left panels, together with ARIs  in the right panels.
When $h=0$, the gaCRP method reduces to the traditional CRP method, which always tends to over-cluster and shows smaller ARI than results by LPML method. 
We see that, when $h$ increases, the estimated number of groups firstly decreases and then increases from histograms in Figure \ref{fig:sim_res_merge} . The reason is that as $h \rightarrow \infty$, the graphical weight $w_{ij}$ for disconnected nodes becomes particularly small, which means only connected nodes can be classified into the same group, therefore leading to the over-cluster problem, as we discussed in Remark \ref{remark.2}.
The group concordance under optimal $h$ selected by LPML generally performs better than those of EM method, with the proportion of selecting true number of groups always greater than 80\%. 


\begin{table}[]
\centering
\caption{RMSE of parameter estimates. For each parameter, the left column comes from scenario 1, and the right column is from scenario 2.}
\scalebox{0.8}{
\begin{tabular}{c|cc|cc|cc|cc|cc}
\toprule
        & \multicolumn{2}{c|}{$\beta_0$} & \multicolumn{2}{c|}{$\beta_1$}  & \multicolumn{2}{c|}{$\beta_2$}  & \multicolumn{2}{c|}{$\gamma$} & \multicolumn{2}{c}{$\sigma^2$}  \\ 
       \midrule
        \multicolumn{11}{c}{\textbf{Example 1}} \\
       \midrule 
        LPML & 0.626 &0.471 & 0.179 &0.377 & 0.057 &0.108 & 0.264 &0.532 & 0.330 &0.624 \\ 
        EM & 1.706 &0.233 & 0.164 &0.167 & 0.132 &0.123 & 0.514 &0.279 & 1.450 &1.466 \\ 
        2-step & 3.774 &0.226 & 0.318 &0.242 & 0.496 &0.258 & 0.717 &0.380 & 1.128 & 1.474\\ 
        \midrule
        \multicolumn{11}{c}{\textbf{Example 2}} \\
       \midrule 
        LPML & 0.722 &0.265 & 0.088 &0.176 & 0.085 &0.135 & 0.404 & 0.429 & 0.450 &0.596 \\ 
        EM & 2.452& 0.317 & 0.329& 0.186 & 0.157& 0.149 & 0.963 & 0.482 & 1.245 & 1.241 \\ 
        2-step & 3.191& 0.370 & 0.329& 0.282 & 0.508& 0.371 & 0.956 & 0.726 & 1.188 & 1.309 \\ 
      \midrule
        \multicolumn{11}{c}{\textbf{Example 3}} \\
       \midrule
        LPML & 0.920& 0.644 & 0.158& 0.180 & 0.093 & 0.089 & 0.460 & 0.318 & 0.443 & 0.246 \\ 
        EM & 1.942 & 1.416 & 0.300& 0.288 & 0.110& 0.120  & 0.748 & 0.573 &1.183 & 0.576 \\ 
        2-step & 2.880 & 1.890 & 0.356  & 0.315 & 0.568 & 0.480 & 0.878 & 0.723 & 0.948 & 0.576 \\ 
   \bottomrule
\end{tabular}
}
\label{table:SBM_simu_res1}
\end{table}

\section{Empirical Case Studies}\label{sec:real_dat}

\begin{figure}[htpb!]
    \includegraphics[width=0.45\textwidth]{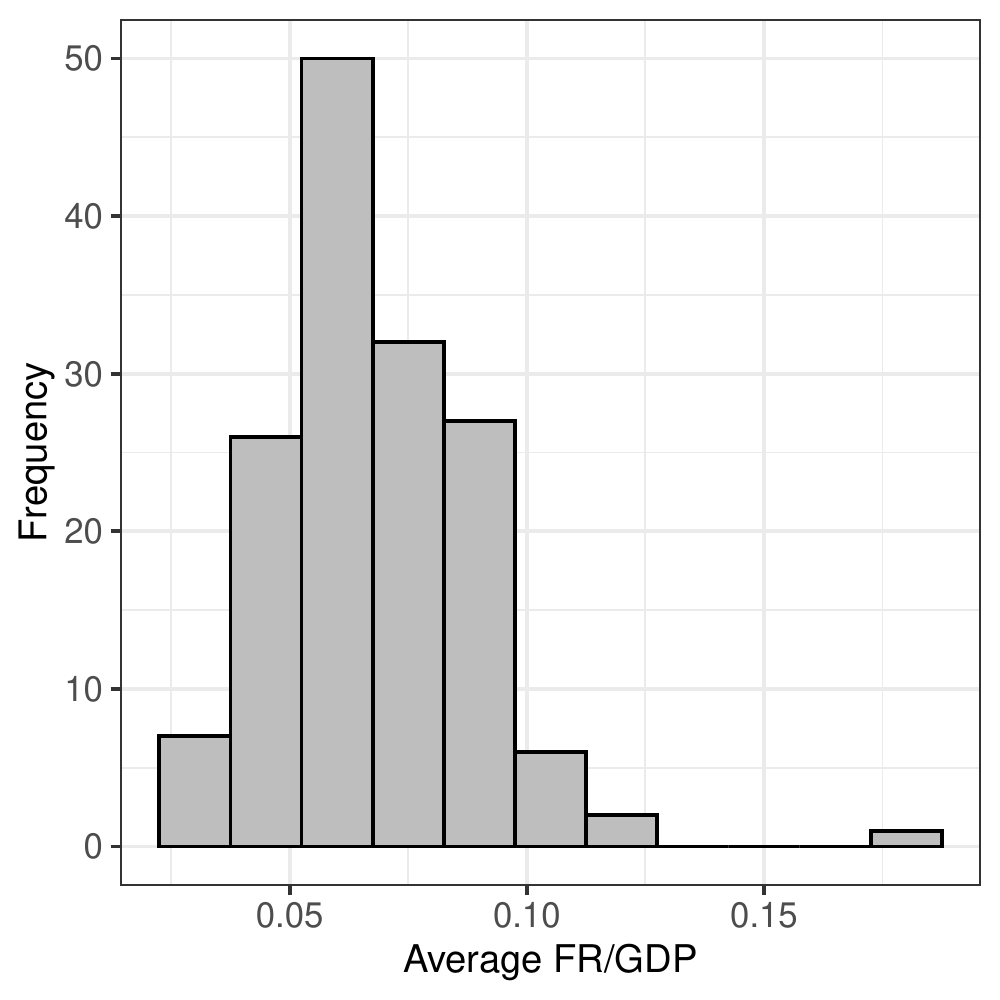}
    \includegraphics[width=0.45\textwidth]{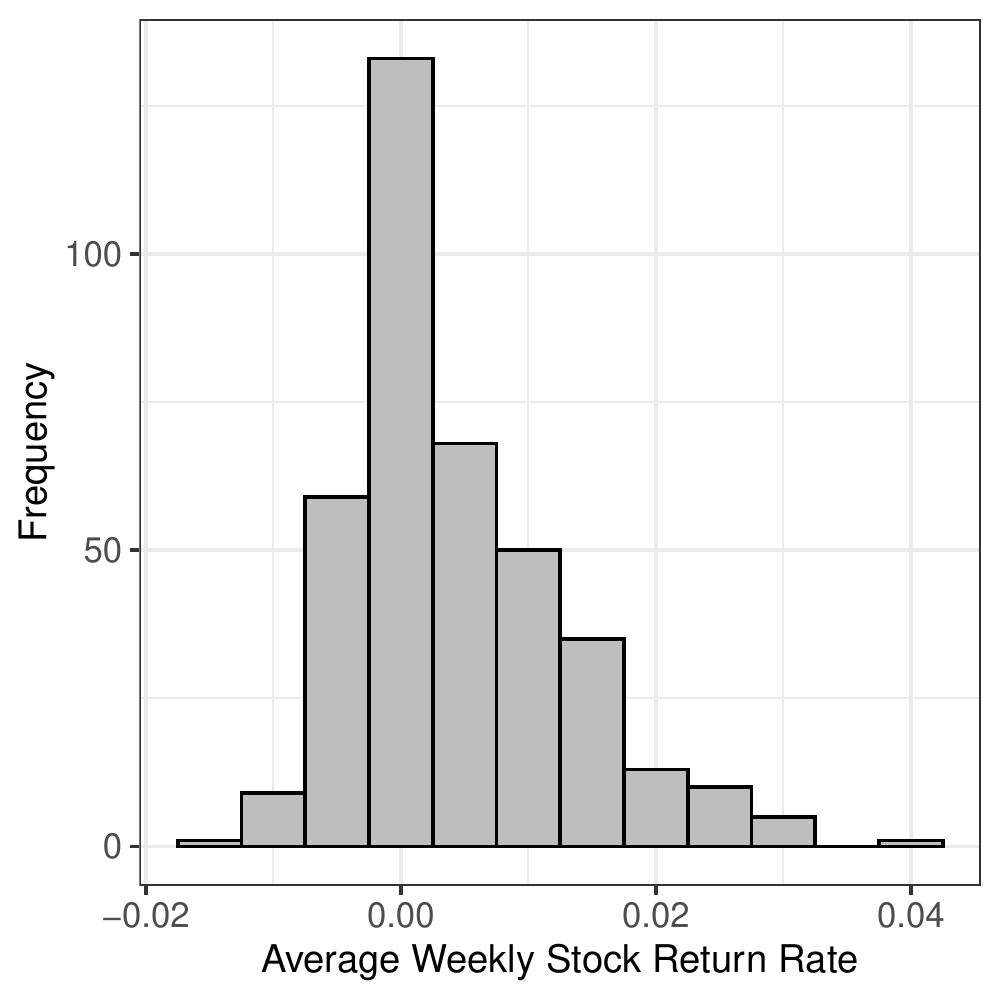}
\caption{\small Left panel: histogram of $N=151$ cities' yearly average fiscal FR/GDP. Right panel: histogram of $N=384$ stocks' weekly average returns.}
\label{fig:real_dsc}
\end{figure}

In this section, we conduct two empirical case studies based on the graphs as Examples 2--3 in Section \ref{sec:simu}. 
For the first real data example, we study the local government economic competition by using city-level fiscal revenue (FR) data. 
For the second real data example, 
we investigate the dynamic patterns of the stocks
traded in the Shanghai and Shenzhen Stock Exchange.
We first conduct a descriptive analysis of the real datasets. We calculate the average ratio of fiscal revenue to GDP from 2002 to 2019 and the average weekly stock return rate during 2019, as shown in Figure \ref{fig:real_dsc}.
The average FR/GDP ratio is 0.068 and the weekly return of the stocks is 0.43\%.

\subsection{China Fiscal Revenue}\label{subsec:real_city}

\subsubsection{Background and Data Description}

We first investigate how the city fiscal revenue
relates to  its spatial adjacent cities and its historical information. 
The model with space–time lags has been widely used on economic panel data research, including analysis about equation systems with time dynamics and spatial spillover effects in regional science \citep{de2012regional,gebremariam2011employment}, fiscal policy with government competition \citep{hauptmeier2012fiscal,allers2011simultaneous},
and many related fields 
\citep{korniotis2010estimating,brown2012they}.
In addition, the latent group structure has been considered in recent macro-economic researches. 
For instance, \cite{bonhomme2015grouped} proposed the time-varying grouped patterns of heterogeneity in linear panel data models, and model the relationship between the degree of democracy and GDP. 
They found the group patterns across countries.
A simple and fast approach was proposed by \cite{liu2020identification} to identify and estimate the unknown group structure in panel data models, which is used to analyze the aggregate production function. They discovered both individual-level and group-level heterogeneity for women's labor force participation.


We collect $N =  151$ city public financial statements from {\it Fiscal Statistics of Cities and Counties in China} during the period 2002-2019. 
The yearbook is published by China Financial and Economic Publishing House, a state-owned press under the supervision of the Ministry of Finance of the People’s Republic of China. 
It collects detailed information on city-level fiscal statistics, such as fiscal revenues and expenditures, fiscal accounting balances, transfer payments, and the fiscally-supported population \citep{yu2016strategic}. 
The response $Y_{it}$ is the fiscal revenue divided by local GDP. 
Following \cite{zhang1998fiscal}, \cite{devereux2007horizontal} and \cite{lv2020fiscal}, 
we consider four covariates, which are POP (population at the end of year), GDP1st (the proportion of primary industry to GDP), SAV (year-end savings of urban and rural residents),
and FOR (actual foreign investment). 
Histogram in the left panel of Figure \ref{fig:real_dsc} shows the average FR/GDP across all cities. 
A right skewed distribution pattern can be observed.

\subsubsection{Model Estimation}

\begin{figure}[htpb!]
    \includegraphics[width=0.5\textwidth]{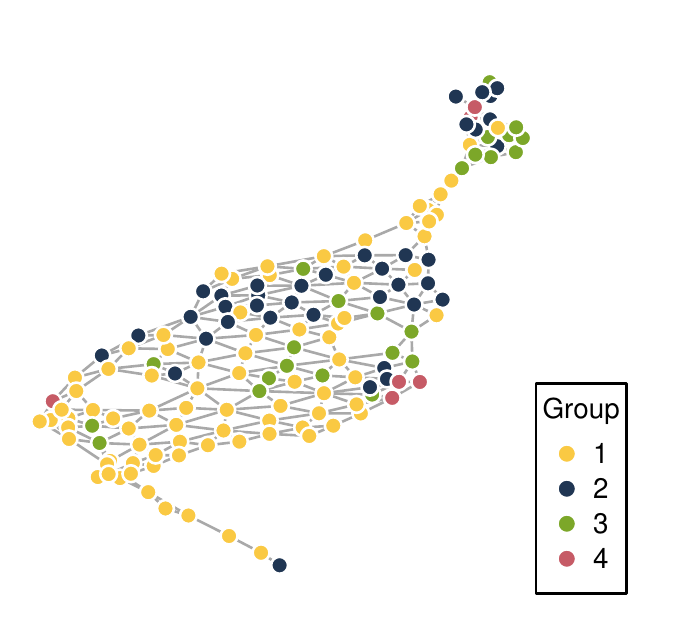}
    \includegraphics[width=0.5\textwidth]{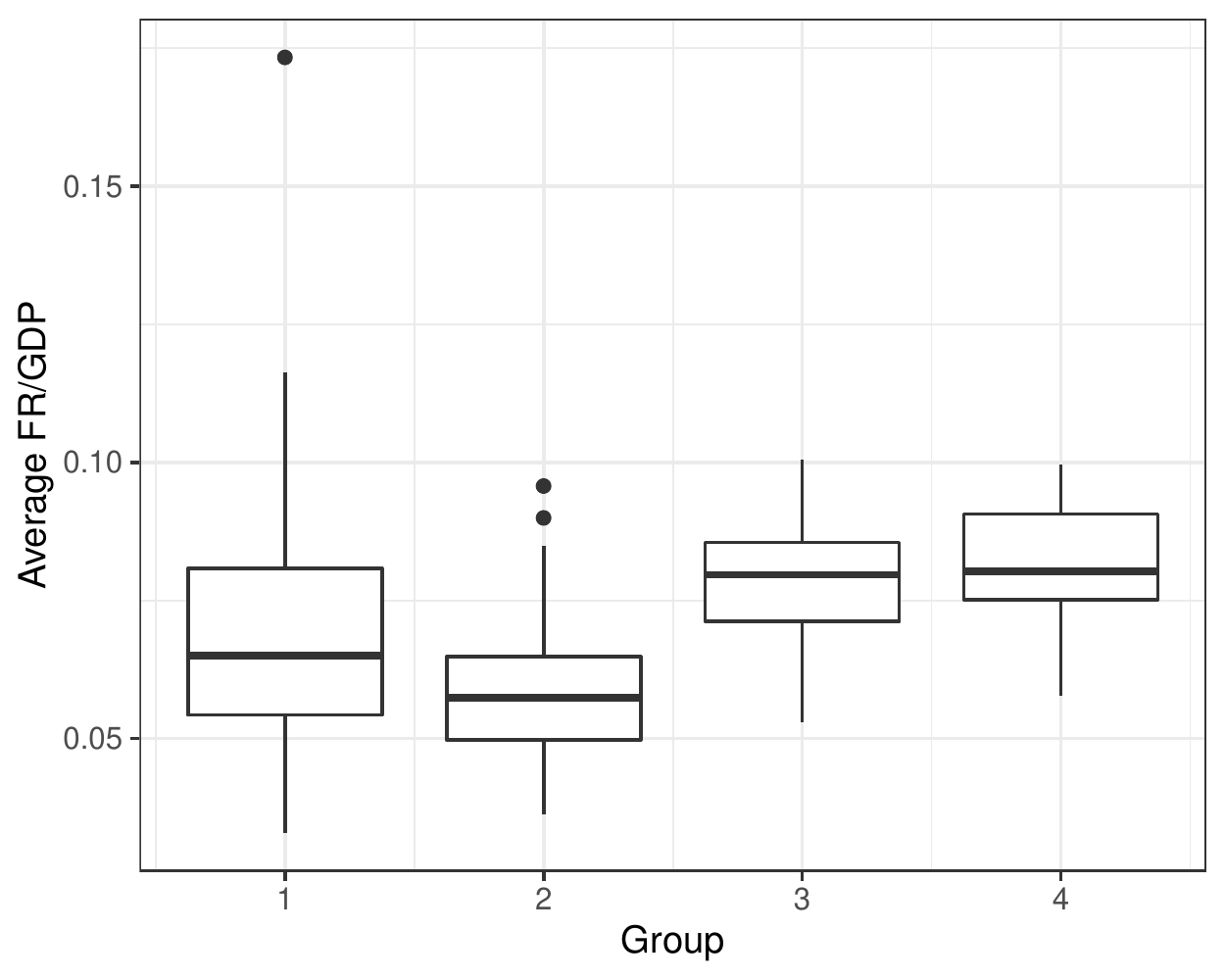}
    \includegraphics[width=0.5\textwidth]{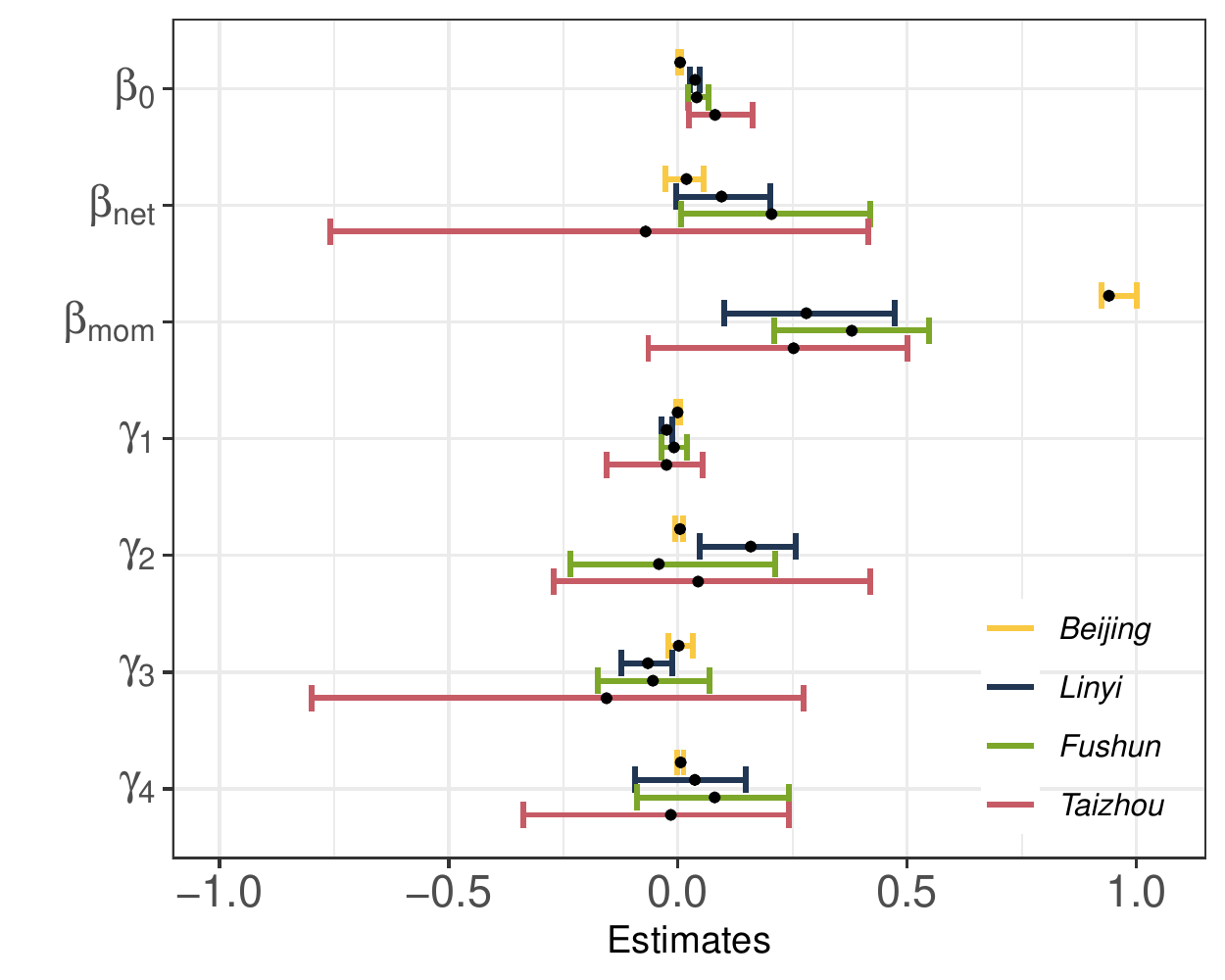}
    \includegraphics[width=0.5\textwidth]{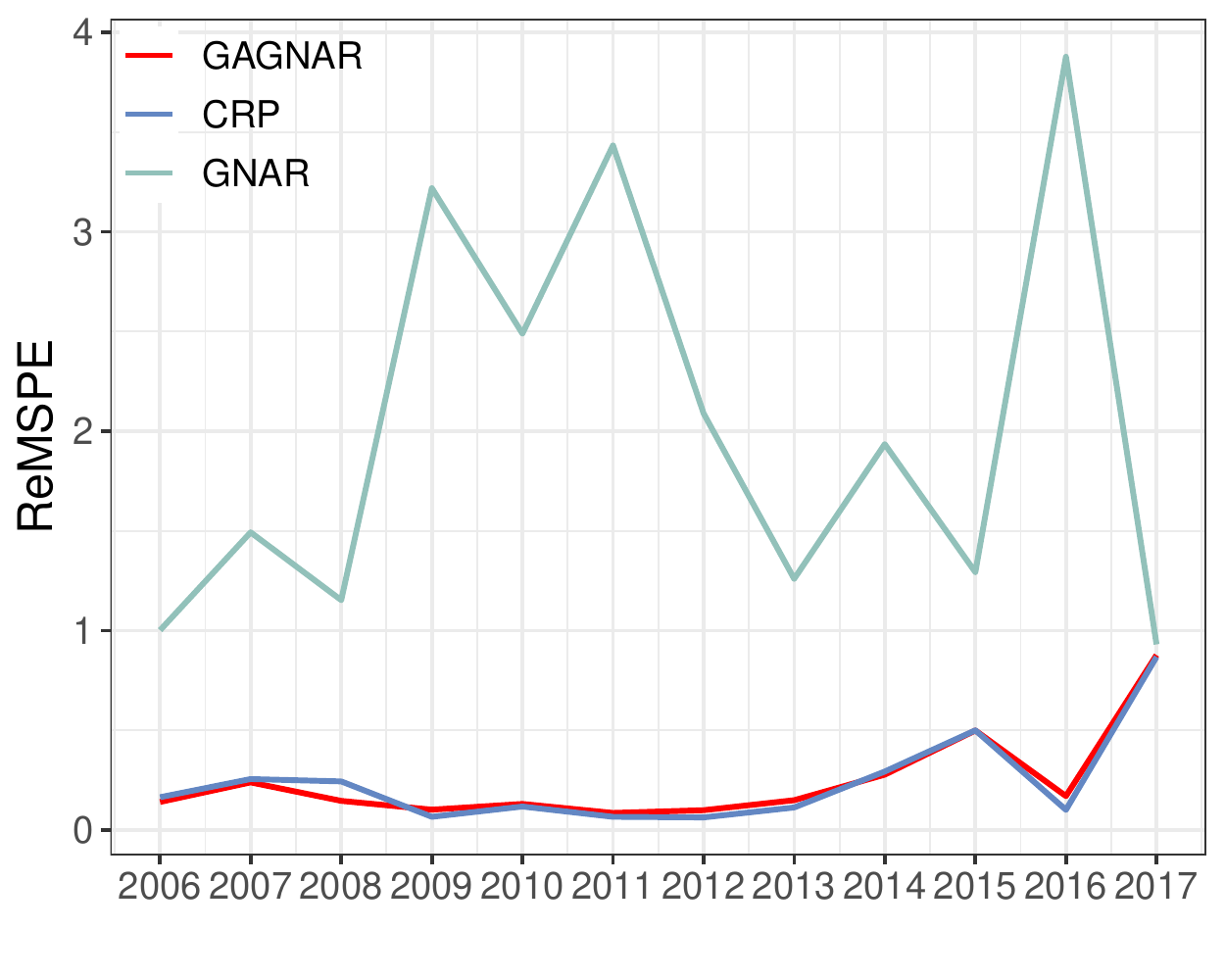}
\caption{\small The top left panel shows the group pattern of cities estimated by GAGNAR under $h=2$. The nodes in group 1--4 are marked in yellow, dark blue, green and pink. The top right panel shows the average FR/GDP of cities corresponding to four estimated groups. Bottom left panel shows the 95\% HPD intervals of four cities from different groups, and the ReMSPE of response in 2006--2017 is displayed in the bottom right panel.}
\label{fig:real_res_city}
\end{figure}

We next implement the GAGNAR model \eqref{eq:hierearchical_model} to the China fiscal revenue data. 
We use the data from 2005 to 2016 as the training set, and make prediction from 2006 to 2017.
The smoothing parameter $h$ is chosen from 
$\{0,0.2,0.4,\cdots,2.0,3.0,4.0,5.0\}$ and
we set $\alpha = 1$. 
For each $h$, we run 1500 MCMC iterations and drop the first 500 as burn-in. The best $h$ selected by LPML is 2, and the estimated number of groups is $K=4$. 
The city group pattern under $h=2$ is displayed in the top left panel of Figure \ref{fig:real_res_city}. 
Cities in the first group are mainly from Guangdong, Jiangxi, Zhejiang and Anhui province, which are located  in the south China.
The second group contains mainly the cities in Henan and Shandong province (east China), with the average FR/GDP at the lowest level as the top right panel of Figure \ref{fig:real_res_city} shows. 
Most cities in Liaoning, Jilin, Heilongjiang provinces and others constitute the third group, which are mostly in the Northeast China. 
Some cities in Jiangsu are in the fourth group, with the highest average FR/GDP values. 
According to the top right panel of Figure \ref{fig:real_res_city}, cities in four groups exhibit different economic features, as the average FR/GDPs among four groups show different levels.

Table \ref{table:est_china_city} reports the estimated parameter for each group under the optimal LPML criterion. 
As one could see, the network effect and momentum effect are different among four groups. 
Specifically, the cities in the second and the third group have positive network effects, which means there might exist imitation economic strategies for those cities with their spatial neighbors. 
On contrary, cities in the first and the last group are negatively related to their neighbors. 
This implies that they tend to take opposite macroeconomic decisions to their adjacent cities.
All of the four groups have positive momentum effects, which shows that the responses are positively related to their historical performances. 
Furthermore, the covariates tend to have different effects on the response. 
For example, the population has negative effect on the fiscal revenue for cities in the first and the second group while its influence is positive for other cities. 
The foreign investment, which represents the openness of a city, negatively affects the fiscal revenue for cities in the fourth group, while it is positive for other cities.
To summarize the posterior distribution of each parameter, the Highest Posterior Density (HPD) \citep{chen1999monte,kruschke2014doing} interval that spans the majority of the posterior distribution is specified.
We calculate the 95\% HPD intervals of parameters of four typical cities (Beijing, Linyi, Fushun and Taizhou) from four groups using the R package {\it HDInterval}. The results are reported in the bottom left panel of Figure \ref{fig:real_res_city}.
Here the HPD interval is calculated for each city specific parameters.
According to the result, one could see that though the coefficients of POP are similar among four groups, the other parameters are quite different across the four cities. 

Lastly, we compare the prediction accuracy of our model with CRP as well as the GNAR model. 
Denote $T_{train}$ and $T_{test}$ as the time length of training set and testing set respectively. 
To evaluate the prediction accuracy, 
we calculate the mean square prediction error $\text{MSPE} = (NT_{test})^{-1} \sum_{t = t_{test,0}}^{t_{test, 0} + T_{test}} \sum_i (\wh{Y}_{it} - Y_{it})^2$
where $\wh{Y}_{it}$ is the predicted response for node $i$ at time point $t$
and $t_{test, 0}$ is the starting time point of the testing set.
Define $\text{MSPE}_0 = (NT_{test})^{-1} \sum_{t = t_{test,0}}^{t_{test, 0}+T_{test}} \sum_i (Y_{it} - \wh{\mu}_{i, train})^2$ as the baseline MSPE, where $\wh{\mu}_{i, train} = T_{train}^{-1} \sum_{t=t_{test,0}- T_{train}}^{t_{test, 0} - 1} Y_{it}$
as the mean response of the $i$th node in the training set.
The relative prediction error is defined as 
\begin{align}
    \text{ReMSPE} = \text{MSPE}_{pred}/\text{MSPE}_0,\label{remspe}
\end{align}
which is used to evaluate the model performance. 
As shown in the bottom right panel of Figure \ref{fig:real_res_city}, when the time length of training data is short, the prediction accuracy of GNAR model is lower and not robust.
The prediction performances of our model and the CRP  model are comparable,
while our model achieves slightly higher accuracy level.

\begin{table}[]
    \centering
    \caption{Parameter estimated by China Fiscal Revenue dataset with optimal $h=2$.}
    \scalebox{0.8}{
    \begin{tabular}{c|c|c|c|c|c}
         \toprule
    \multicolumn{2}{c|}{} & Group 1& Group 2& Group 3 & Group 4\\
    \midrule
    \multicolumn{2}{c|}{$N_k$} & 83 & 38 & 25 & 5 \\
    \midrule
    \multicolumn{2}{c|}{$\wh{\beta}_{0k}$} &0.003 &0.037 &0.038 & 0.047\\
    \midrule
    $\wh{\beta}_{1k}$ & network effect &-0.011 &0.101 & 0.212&-0.179 \\
    $\wh{\beta}_{2k}$ & momentum effect &0.996 &0.229 &0.345 &0.177 \\
    \midrule
    \multirow{4}*{$\wh{\gamma}_k$} & POP &-0.001 &-0.025 &0.022 &0.113 \\
    & GDP1st &-0.001 &0.175 &-0.128 &0.186 \\
    & SAV &0.008 & -0.066&0.001&-0.074\\
    & FOR &0.008 &0.071 &0.078 &-0.383 \\
    \midrule
    \multicolumn{2}{c|}{$\wh{\sigma}_k^2$} &0.0001&0.0003&0.0006&0.0078\\
    \bottomrule
    \end{tabular}
    }
    \label{table:est_china_city}
\end{table}

{
\subsection{Stock Return Analysis with Shareholder Network}\label{subsec:real_stock}

\subsubsection{Background and Data Description}

The network information has been taken into consideration as an important factor recently in analyzing financial time series.
For instance, \cite{zhu2019network} investigate the financial contagion phenomenon in the stock market with a common shareholder network. To improve  predictability of stocks return, \cite{leung2017network} define a search-based cluster using co-search network of stocks on Yahoo.com.
In this study, we aim to analyze the stock returns 
using a common shareholder network.
Specifically, we collect $N=384$ stocks traded in the Shanghai and the Shenzhen Stock Exchange.
The response $Y_{it}$ is the weekly stock return for 52 weeks in the year of 2019.
Motivated by \cite{fama2015five}, we consider six covariates related to corporations' fundamentals.
They are, {\sc SIZE} (the logarithm of market value),
{\sc BM} (book to market value),
{\sc PR} (yearly incremental profit ratio),
{\sc AR} (yearly incremental asset ratio),
{\sc LEV} (leverage ratio),
and {\sc CASH} (cash flow). 
The histogram in the right panel of Figure \ref{fig:real_dsc} shows the average return during the whole year, calculated by $T^{-1} \sum_t Y_{it}$. 
The mean weekly return rate of all stocks is given by 0.43\%.

\subsubsection{Model Estimation}

To characterize the dynamics of the stock return, the GAGNAR model is implemented to investigate the group patterns among the stocks.
We use data in the 52 traded weeks, with $T_{train}=40$ weeks used as the training set and the remaining $T_{test}=12$ as the test set. The smoothing parameters are selected from $h\in \{0,0.2,0.4,0.6,\dots, 3.0,4.0,5.0\}$
For each $h$, 1500 MCMC iterations are run, and we drop the first 500 as burn-in. 
The best $h$ selected by LPML is 0.6, under which the number of groups is estimated as $K=5$. 
The group pattern is shown in the top left panel of Figure \ref{fig:real_res_stk}. 
We then look into the industry information of the five groups.
Stocks in the first group are mainly from the real estate industry. The second group has the majority of stocks in information technology industry. Stocks from electrical machinery and equipment manufacturing industry constitute the major part of the third group, while  stocks from the production and supply of electricity, steam and hot water form the fourth group. 
Lastly, the chemical related industries as well as civil engineering construction are included in 
the fifth group.
In addition, the top right panel of Figure \ref{fig:real_res_stk} shows that
the stock return levels are diversified in the five groups.
Particularly, the third group achieves the highest average weekly return rate, while stock returns of the first and fourth groups are relatively low.

The parameter estimation is shown in Table \ref{table:est_stock_ret}. As presented in the result, the network effect and momentum effect are different among five groups.
Specifically, the stocks in the second group are positively related to their connected stocks, while the stocks in the rest four group behave oppositely. 
The opposite network effects are also observed in empirical studies in financial stock market \citep{borgatti2009network, peng2013research, chen2019can}.
The momentum effects among the first four groups are negative, which shows that they are conversely related to their historical return rate. 
The last group, on contrary, is positively related to its historical behaviors.
Furthermore, we observe that the covariates have different effects on the response. 
For example, the market value (SIZE) of stocks in the first, second and the fourth group
positively contribute to the weekly traded return.
The cash flow (CASH) of the stocks in the second group is positively related to the return rate, while it is not the case for other four groups.
Next, the 95\% HPD intervals are depicted in the bottom left panel of Figure \ref{fig:real_res_stk}, from which one could see that there are obvious dissimilarity in the network effect and momentum effect among stocks in different groups. Though there is slight difference in the covariates coefficients among five groups, our proposed model could capture the imperceptible distinction for different nodes.

Lastly, to evaluate the out-sample prediction performance of the GAGNAR model, we also use ReMSPE \eqref{remspe} to compare the prediction accuracy of GAGNAR with CRP and the GNAR model. The results are shown in bottom right  panel of Figure \ref{fig:real_res_stk}. 
As one could observe, the GAGNAR model achieves slightly higher and more robust prediction accuracy than other competing methods.
This illustrates the usefulness of the proposed methodology.

\begin{table}[]
    \centering
    \caption{Parameter estimated ($\times $10) by Stock Return dataset with optimal $h=0.6$.} 
    \scalebox{0.8}{
    \begin{tabular}{c|c|c|c|c|c|c}
    \toprule
    \multicolumn{2}{c|}{} & Group 1& Group 2& Group 3 & Group 4&Group 5\\
    \midrule    
    \multicolumn{2}{c|}{$N_k$} & 92 & 94 & 124 & 71 & 3 \\ 
    \midrule    
    \multicolumn{2}{c|}{$\wh{\beta}_{0k}$} & 0.008 & 0.154 & 0.029 & -0.007 & 6.164\\
    \midrule    
    $\wh{\beta}_{1k}$ & network effect & -0.313 & 0.343 & -0.678 & -0.172 & -7.776 \\
    $\wh{\beta}_{2k}$ & momentum effect & -1.066 & -0.541 & -0.692 & -1.032 & 1.795 \\
    \midrule
    \multirow{4}*{$\wh{\gamma}_k$} & SIZE & 0.006 & 0.009 & -0.006 & 0.001 & -4.968\\
    & BM & -0.003 & -0.01 & 0.03 & -0.012 & 9.136\\
    & PR & -0.006 & 0.007 & -0.014 & 0.011 & -2.988\\
    & AR & -0.016 & 0.077 & 0.008 & 0.001 & -4.028\\
    & LEV & 0.011 & 0.033 & -0.018 & 0.003 & 11.44 \\
    & CASH & -0.001 & 0.023 & -0.002 & -0.009 & -13.783\\
    \midrule
    \multicolumn{2}{c|}{$\wh{\sigma}_k^2$} & 0.025 & 0.074 & 0.037 & 0.012 & 0.228\\
    \bottomrule
    \end{tabular}
    }
    \label{table:est_stock_ret}
\end{table}

\begin{figure}[htpb!]
    \includegraphics[width=0.5\textwidth]{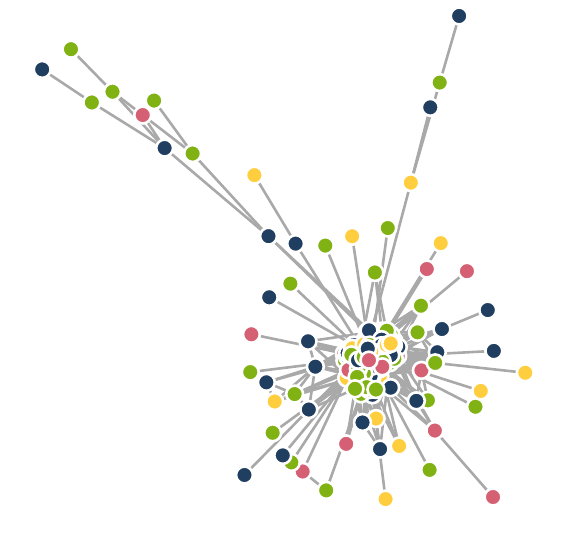}
    \includegraphics[width=0.5\textwidth]{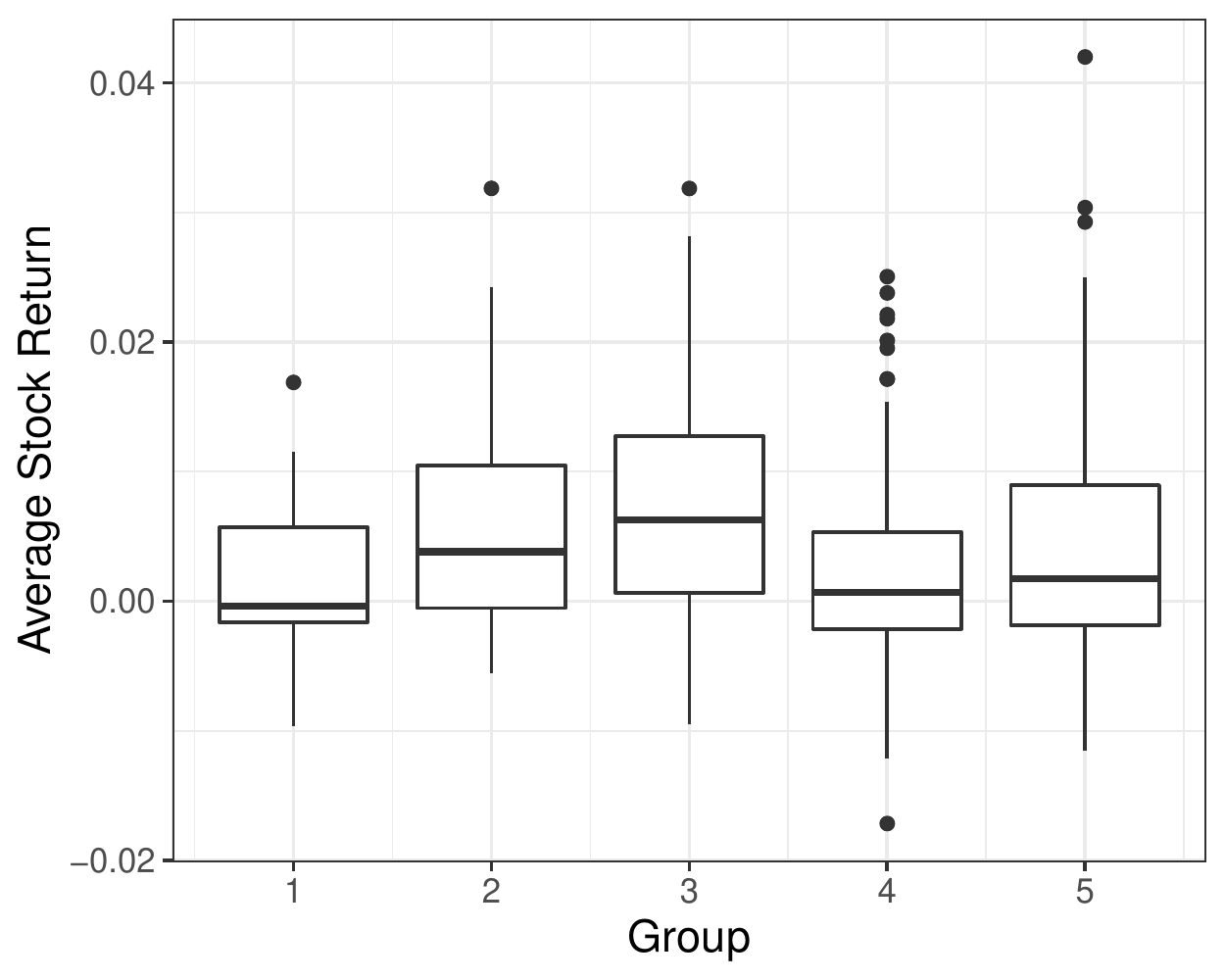}
    \includegraphics[width=0.5\textwidth]{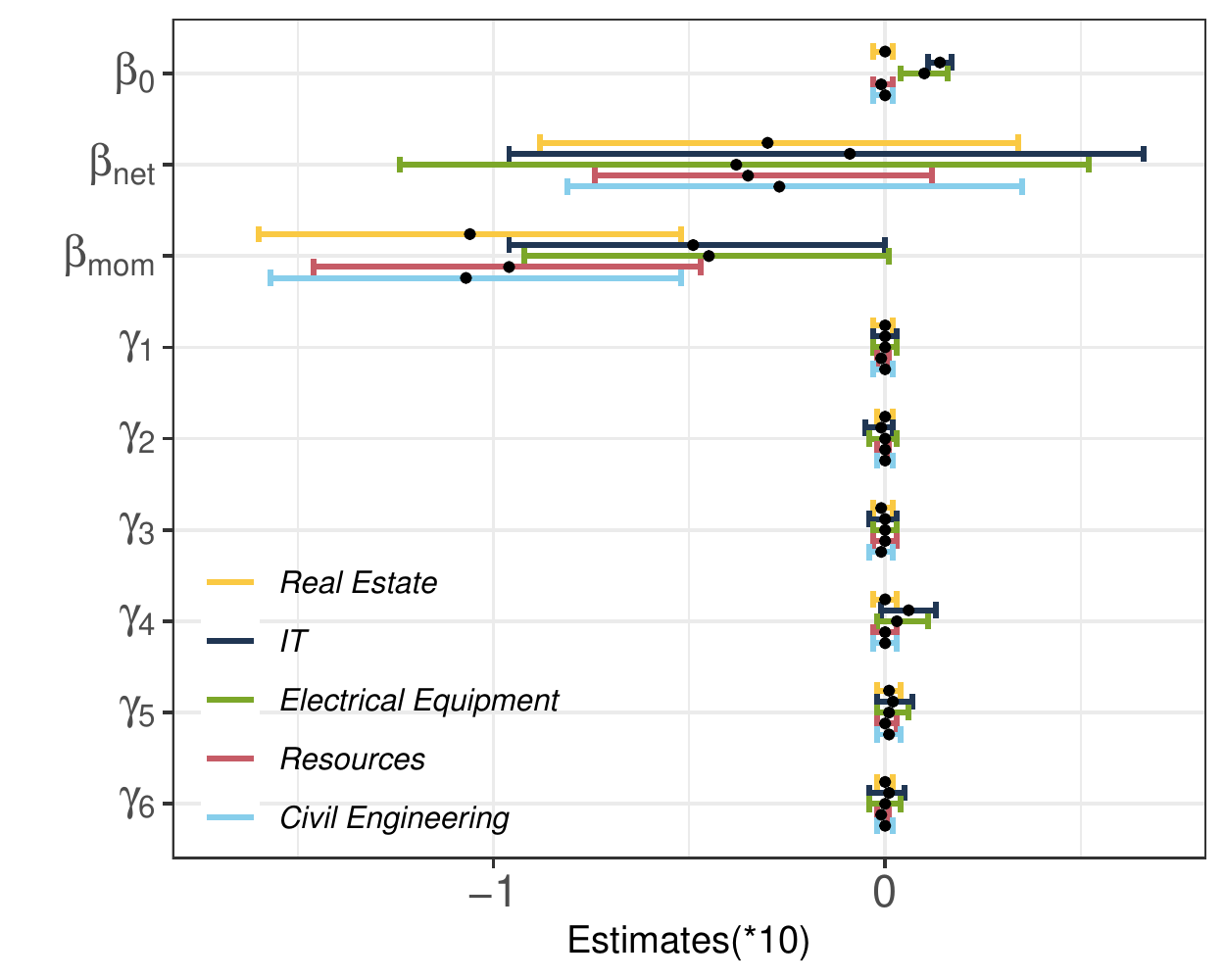}
    \includegraphics[width=0.5\textwidth]{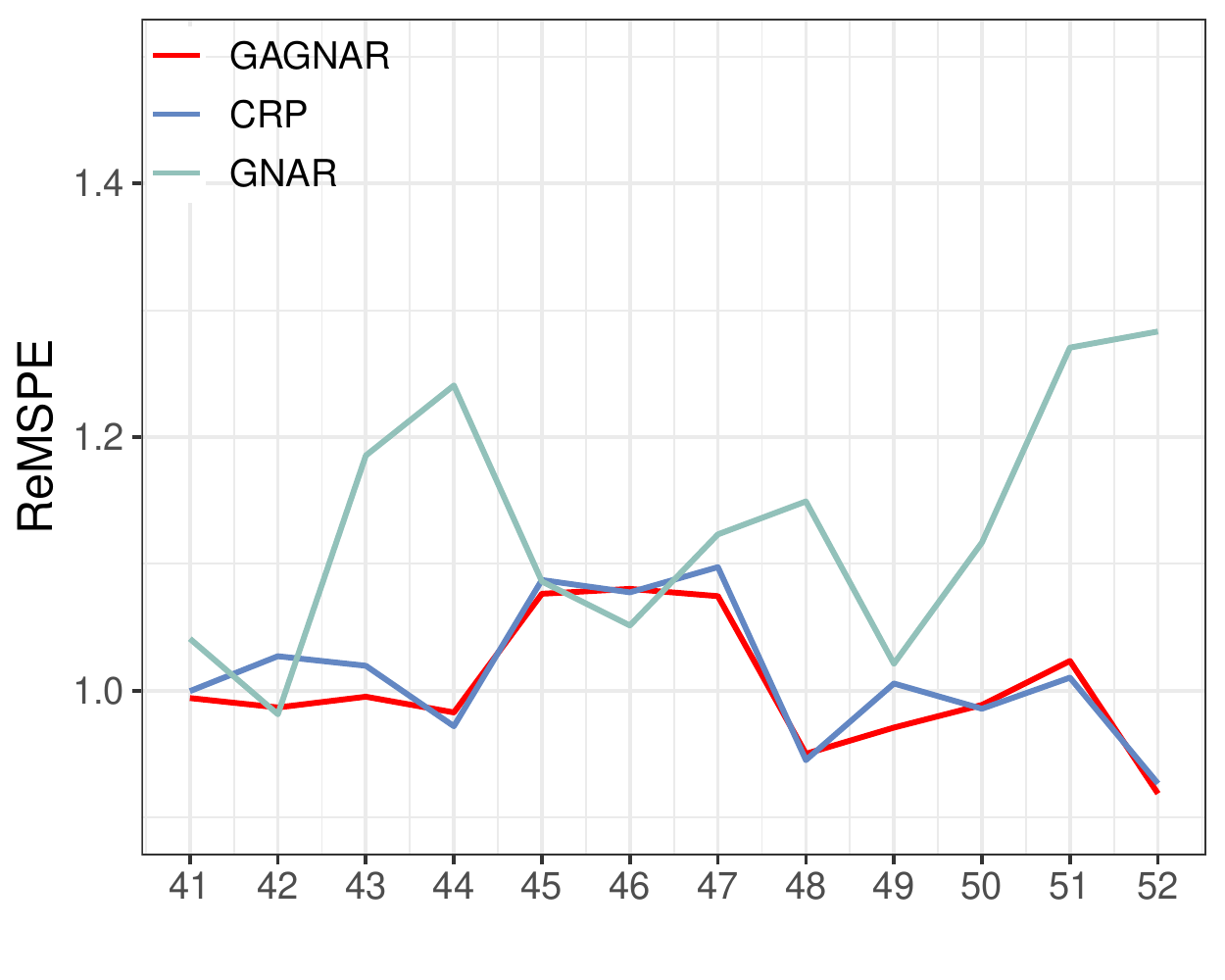}
\caption{\small The top left panel shows the group pattern of stocks estimated by GAGNAR under $h=0.6$. 
The nodes in group 1--5 are marked in yellow, dark blue, green, pink, and skyblue.
The top right panel shows the average return rate of stocks corresponding to five estimated groups. Bottom left panel shows the 95\% HPD intervals of five stocks, which are the first stock of the five groups sorted by the stock code respectively. The ReMSPE in week 41--52 is displayed in the bottom right panel.}
\label{fig:real_res_stk}
\end{figure}

}

\section{Discussion}\label{sec:discussion}

In this paper, we propose a novel Bayesian nonparametric grouping approach for learning the heterogeneity of dynamic response for network data. 
The proposed method is able to conduct model estimation, estimate the number of groups, and infer group configurations simultaneously.
Building upon the gaCRP framework, we develop a collapsed Gibbs sampler for efficient Bayesian inference.
Specifically we adopt Dahl's method for post MCMC inference and introduced LPML for smoothing parameter selection. 
Numerical results have confirmed that the proposed method is able to simultaneously infer the number of groups and the group-wise parameters with high accuracy. 
Comparing to the traditional techniques such as EM algorithm and two-step approach in \citet{zhu2018grouped}, the proposed method is able to improve the grouping performance, especially when grouping configurations contain certain graphical information.
Lastly, we illustrate the usefulness of the proposed method by studying two real data examples, including 
China fiscal revenue analysis and stock return prediction.
The results indicate that incorporating graphical information on grouping process will improve the interpretability and  prediction power of the methodology.

A few topics beyond the scope of this paper are worth further investigation. 
A natural extension for this paper is for large scale networks.
Designing an efficient algorithm for massive datasets with lower computational complexity devotes an interesting future work. 
Two promising solutions for tackling this challenge are discovering low rank structure and imposing sparsity. 
Second, the proposed method is restricted to continuous responses and thus can be extended to binary or counted data.
Finally, incorporating the GNAR model with the stochastic block model (SBM) will be another interesting topic for future study.

\bibliographystyle{chicago}

\end{document}